\documentclass[11pt]{emulateapj}
\usepackage{natbib}
\usepackage{graphicx}
\usepackage{amsmath}
\usepackage{amssymb}

\def\Wm2{W/m$^2$}
\def\Wpm2sr{Wm$^{-2}sr^{-1}$}
\def\deg{$^\circ$ }
\def\degx{$^\circ$}

\begin{document}
\title{Uranus at Equinox: cloud morphology and
dynamics\footnotemark[\dag]}
\author{L.~A. Sromovsky\altaffilmark{1},
  P.~M. Fry\altaffilmark{1},
  H.~B. Hammel\altaffilmark{2}$^,$\altaffilmark{3}, W.\ M. Ahue\altaffilmark{1} I. de Pater\altaffilmark{4},
   K.~A. Rages\altaffilmark{5}, M.~R. Showalter\altaffilmark{5}, and M.~A. van Dam\altaffilmark{6}}
\altaffiltext{1}{University of Wisconsin - Madison, Madison WI 53706}
\altaffiltext{2}{AURA, 1212 New York Ave. NW, Suite 450, Washington, DC 20005, USA}
\altaffiltext{3}{Space Science Institute, Boulder, CO 80303, USA}
\altaffiltext{4}{University of California, Berkeley, CA 94720, USA}
\altaffiltext{5}{SETI Institute, Mountain View, CA 94043, USA}
\altaffiltext{6}{W.~M. Keck Observatory, Kamuela, HI 96743, USA }
\altaffiltext{\dag}{Based in part on observations with the NASA/ESA Hubble Space
Telescope obtained at the Space Telescope Science Institute, which is
operated by the Association of Universities for Research in Astronomy,
Incorporated under NASA Contract NAS5-26555.} 

\slugcomment{Journal reference: Icarus 203 (2009) 265-286.}
\begin{abstract}

As the 7 December 2007 equinox of Uranus approached, collaboration
between ring and atmosphere observers in the summer and fall of 2007
produced a substantial collection of groundbased observations using
the 10-m Keck telescope with adaptive optics and space-based
observations with the Hubble Space Telescope. Both near-infrared and
visible-wavelength imaging and spatially resolved near-infrared
spectroscopic observations were obtained. We used observations
spanning the period from 7 June 2007 through 9 September 2007 to
identify and track cloud features, determine atmospheric motions,
characterize cloud morphology and dynamics, and define changes in
atmospheric band structure.  Atmospheric motions were obtained over a
wider range of latitudes than previously was possible, extending to
73\deg N, and for 28 cloud features we obtained extremely high
wind-speed accuracy through extended tracking times. We confirmed
the existence of the suspected northern hemisphere prograde jet,
locating its peak near 58\deg N.  The new results
confirm a small N-S asymmetry in the zonal wind profile, and the lack
of any change in the southern hemisphere between 1986 (near solstice)
and 2007 (near equinox) suggests that the asymmetry may be permanent
rather than seasonally reversing.  In the 2007 images we found two
prominent groups of discrete cloud features with very long
lifetimes. The one near 30\deg S has departed from its previous
oscillatory motion and started a significant northward drift,
accompanied by substantial morphological changes.  The complex of
features near 30\deg N remained at a nearly fixed latitude, while
exhibiting some characteristics of a dark spot accompanied by bright
companion features. Smaller and less stable features were used to
track cloud motions at other latitudes, some of which lasted over many
planet rotations, though many could not be tracked beyond a single
transit.  A bright band has developed near 45\deg N, while the bright
band near 45\deg S has begun to decline, both events in agreement
with the idea that the asymmetric band structure of Uranus is a
delayed response to solar forcing, but with a surprisingly short delay
of only a few years.

\end{abstract}


\maketitle
\shortauthors{Sromovsky et al.} 
\shorttitle{Uranus at equinox: Cloud morphology and dynamics}

\section{Introduction}

Uranus reached equinox on 7 December 2007, for the first time in 42
years, and for the first time ever when modern high resolution
instruments were capable of detailed characterization of the
event. Uranus' appearance in HST and Keck imagery at near-IR
wavelengths revealed a richer treasure-trove of atmospheric
information than the bland close-up Voyager imaging in 1986.  Recent
pre-equinox observations recorded a surprising level of activity,
unusual cloud dynamics, and a hemispherical asymmetry in cloud bands,
discrete cloud structure, and atmospheric motions.  A wide range of
discrete cloud lifetimes was discovered, ranging from less than an
hour to possibly decades, and a north-south asymmetry in the zonal
wind was characterized \citep{Sro2005dyn}.  A recent dramatic event
was the rise and fall of Uranus' brightest cloud feature, which peaked
in 2005, and faded dramatically thereafter \citep{Sro2007bright}. A
large and very long lived storm system (S34) was discovered
\citep{Sro2005dyn}. Some evidence suggests that it existed as early as
the 1986 Voyager encounter, but there is little doubt that it existed
from at least 2000 through at least 2005, during which time it slowly
oscillated in latitude between 32\deg S and 36\deg S and exhibited a
superimposed inertial oscillation with a period near 0.7 days. In 2004
it suddenly became visible in K$'$ images \citep{Hammel2005newact},
indicating an increased cloud altitude. Cycle 15 HST observations in
2006 captured the first dark spot in Uranus' northern hemisphere
\citep{Hammel2008spot}.  This roughly elliptical spot is similar to
several dark spots observed on Neptune, even exhibiting what appears
to be a bright companion cloud in a pre-discovery image.

New ground-based images we obtained from the Keck telescope in 2007
(beginning on 7 June and extending through 9 September) provide a new
abundance of discrete cloud features that we use to gain a better
understanding of Uranus' atmospheric dynamics.  We also obtained
imagery from the Hubble Space Telescope Wide Field Planetary Camera 2
in July and August of 2007.  Here we discuss cloud morphologies, motions, and
evolution, and include in the analysis results from Voyager and prior
HST observations.  We extend the zonal wind observations up to 73\deg
N, and further establish the north-south asymmetry of Uranus' zonal
circulation. We were able to determine highly accurate drift rates for
28 cloud features by tracking them over long time periods, and find
little change in the measured circulation between 1986 and 2007.

\section{Observations}

Our primary data are Keck and HST observations obtained during 2007. We
also made use of a limited number of HST archival images from 2004-2006.

\subsection{Recent Keck and HST Observations}

\begin{figure*}[!htb]\centering
\includegraphics[width=6in]{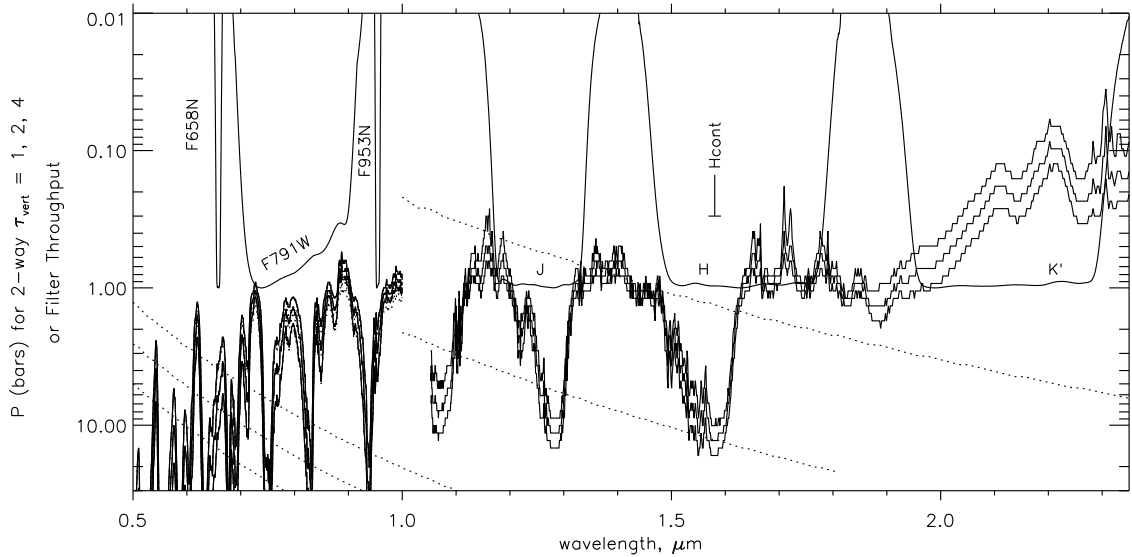}
\caption{Penetration of sunlight into the atmosphere of Uranus vs
wavelength.  Solid curves are shown for two-way vertical optical
depths of 1, 2, and 4 from CH$_4$ and H$_2$ absorption, assuming a 2.26\% CH$_4$ mixing
ratio. Dotted curves
show the same optical depths for Rayleigh scattering, except for
$\lambda > 1 \mu$m, where curves are shown for Rayleigh optical depths
of 0.1 and 0.01.  HST/WFPC2, HST/ACS, and Keck/NIRC2 filters are shown as system
throughput curves, normalized to unity at their peaks. See Table
\ref{Tbl:filtinfo} for additional filter information.}
\label{Fig:specpendepth}
\end{figure*}

Using the NIRC2 camera on the Keck II telescope, suitable atmospheric
 observations of Uranus were obtained on dates given in Table\
 \ref{Tbl:keckobs}, where we also list basic observing conditions and
 PIs for each observing run. The ring-plane crossing was on 16 August, when bad
 weather severely limited observations from the Keck telescope.  
Most of our images were made with broadband J, H, and K$'$ filters using
 the NIRC2 Narrow Camera.  A few were made with narrow-band filters
 for better discrimination of vertical structure.  After geometric
 correction, the angular scale of the NIRC2 narrow-angle camera is
 0.009942$''$/pixel (NIRC2 General Specifications web page:
 http://alamoana.keck.hawaii.edu/inst/nirc2/ \mbox{genspecs.html}). This
 provides 350 (June) to 369 (September) pixels across the
 planet. Typical resolution in the Keck II AO images of Uranus was
 0.06$''$, providing 58-61 resolution elements across the disk.  The
 penetration depth of these observations into the atmosphere of Uranus
 is indicated in Fig.\ \ref{Fig:specpendepth}.  Only particles
 extending to pressures less than 1 bar can be seen in K$'$ images of
 Uranus, while J and H images can see to the 10-bar level in a clear
 atmosphere.  The main variations visible in J and H bands appear to
 occur in the 2-5 bar region \citep{Sro2007struc,Sro2008grism}, well
 below the expected methane condensation level near 1.2 bars.

\begin{table*}\centering
\caption{Uranus Keck II observation summary.} 
\vspace{0.15in}
\begin{tabular}{c | c  r  r  c c l}
   Date   & Temporal          & Sub Obs.& Sub Solar& Uranus& Phase &\\
 (mm/dd/yyyy)   & Coverage (h, UT) & Lat.($\phi_\mathrm{pg}$,\degx) & Lat. ($\phi_\mathrm{pg}$, \degx) & Diam. (as) & Angle (\degx) & PI\\
\hline
06/07/2007 & 14:19-15:23 & 0.97  &  -2.03  &3.504 &2.893    & van Dam\\
07/26/2007 & 10:44-15:11 & 0.66  &  -1.49  &3.639 &2.076    & de Pater\\
07/27/2007 & 10:37-15:34 & 0.64  &  -1.48  &3.641 &2.041    & Hammel\\
07/28/2007 & 12:07-15:12 & 0.61  &  -1.47  &3.643 &2.005    & Hammel\\
07/30/2007 & 11:24-15:28 & 0.56  &  -1.45  &3.648 &1.932    & Sromovsky\\
07/31/2007 & 10:34-15:22 & 0.53  &  -1.43  &3.650 &1.894    & Sromovsky\\
08/08/2007 & 11:15-15:06 & 0.29  &  -1.35  &3.665 &1.574    & de Pater\\
08/09/2007 & 10:53-15:20 & 0.25  &  -1.33  &3.666 &1.531    & de Pater\\
08/19/2007 & 10:30-14:51 &-0.10  &  -1.22  &3.680 &1.083    & Sromovsky\\
08/20/2007 & 10:27-14:49 &-0.14  &  -1.21  &3.682 &1.036    & Sromovsky\\
09/07/2007 & 10:47-13:26 &-0.86  &  -1.01  &3.693 &0.153    & Hammel\\
09/09/2007 & 8:13-13:21  &-0.94  &  -0.99  &3.693 &0.061    & Showalter\\
\hline
\end{tabular}\label{Tbl:keckobs}
\end{table*}

\begin{table*}\centering
\caption{Uranus HST observation summary.} 
\vspace{0.15in}
\begin{tabular}{c| c  r  r  c  c l}
      Date      & Temporal       & Sub Obs.& Sub Solar& Uranus& Phase &\\
(mm/dd/yyy) & Coverage (h UT)& Lat. ($\phi_\mathrm{pg}$, \degx) & Lat. ($\phi_\mathrm{pg}$, \degx) & Diam. (as) & Angle (\degx) & PI\\
\hline
07/28/2007 & 2:48-11:25    & 0.61  &  -1.47  &3.643 &2.005    & Sromovsky\\
07/29/2007 & 2:47-11:23    & 0.59  &  -1.46  &3.645 &1.969    & Sromovsky\\
08/17/2007 & 7:14-7:29   &-0.03  &  -1.25  &3.678 &1.176    & Rages\\
08/19/2007 & 23:16-23:31 &-0.10  &  -1.22  &3.680 &1.083    & Rages\\
08/27/2007 & 6:59-8:53   &-0.41  &  -1.13  &3.688 &0.699    & Rages\\
\hline
\end{tabular}\label{Tbl:hstobs}
\end{table*}

\begin{table}\centering
\caption{Characteristics for NIRC2, WFPC2, and ACS filter bands we used.}
\label{Tbl:filtinfo}
\begin{tabular}{c c c c }
Instrument &  Filter &Central   $\lambda$, $\mu$m &Bandpass$^a$, $\mu$m \\ 
\hline
NIRC2$^b$ & J        & 1.248  & 0.163 \\ 
NIRC2$^b$ & H        & 1.633  & 0.296  \\ 
NIRC2$^b$ & K$'$       & 2.124  & 0.351 \\
NIRC2$^b$ & Hcont    & 1.5804 & 0.0232 \\
  WFPC2$^c$ & 791W  & 0.790 & 0.130  \\
  WFPC2$^c$ & F953N & 0.954 & 0.005 \\
      ACS$^d$   & F658N & 0.658 & 0.008 \\
\hline
\end{tabular}\par
\parbox{2.9in}{\noindent $^a$effective bandpass for WFPC2, otherwise full \newline 
 $^{  }$ width at half maximum.
\newline $^b$www.alamoana.keck/hawaii.edu/inst/nirc2/filters.html.\newline
$^c$WFPC2 Instrument Handbook v 9.0, Oct. 2004.\newline
$^d$ACS Instrument Handbook v 5.0, Oct. 2004.}
\end{table}

Our HST atmospheric observations in 2007 were gathered by a GO program (11118) and a SNAP program (11156)
with observing conditions and PIs given in Table\ \ref{Tbl:hstobs}.
 Their corresponding filter characteristics
are also given in Table \ref{Tbl:filtinfo} and Fig.\ \ref{Fig:specpendepth}.

\subsection{Image Processing and Navigation}\hyphenation{forReDoc}

HST image processing and navigation proceeded along the lines
described by \cite{Sro2007struc}.  Our Keck image processing also followed
\cite{Sro2007struc} except that we used an updated geometric
distortion correction code provided by Brian Cameron of the California Institute
of Technology
(see www2.keck.hawaii.edu/inst/nirc2/forReDoc/post\_ observing/dewarp/).
We made use of the SPICELIB toolkit
\citep{Acton1996} to generate ephemeris information concerning the
orientation of the planet's pole vector, the range to the planet, and
the sub latitude and sub longitude of the observer (the point at which a
vector from the planet center to the observer intersects the surface).
We used standard 1-bar radii of $r_{\mathrm{eq}}$ = 25,559 km and
$r_{\mathrm{pol}}$ = 24,973 km and a longitude system based on a
17.24-h period \citep{Seidelmann2002}.  
We determined planet
center coordinates by fitting a projected planet limb to maximum gradient
limb points.
The combined effect of navigation errors and other cloud tracking
errors is approximately 0.6 narrow angle camera pixels RMS, which is estimated
from the image-to-image scatter found in the plots of discrete cloud
position versus time. In describing positions we use both
planetocentric latitude ($\phi$), which is the angle above the
equatorial plane measured from the planet center, and planetographic
latitude ($\phi_{\mathrm{pg}}=\tan^{-1}[(r_{\mathrm{eq}}/r_{\mathrm{pol}})^2
\tan\phi]$), which is the angle between the local
normal and the equatorial plane. 

\section{Atmospheric Circulation Results}

\subsection{Overview of cloud features in 2007 Keck II images}

\begin{figure*}\centering
\includegraphics[width=5in]{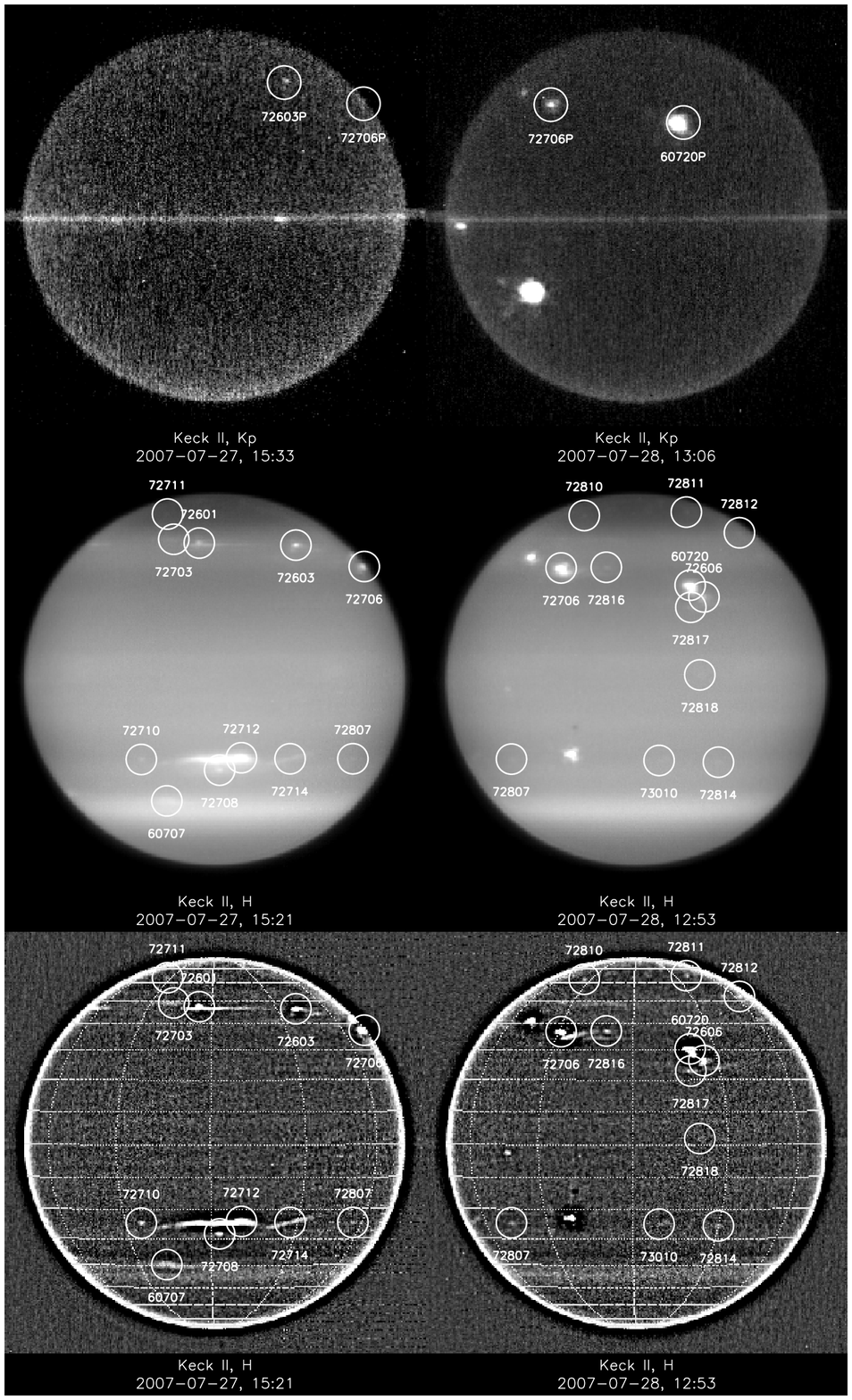}\par
\caption{North-up NIRC2 images of Uranus in K$'$(top) and H filters
made on 27 and 28 July 2007 (the bottom row is high-pass filtered).
Wind measurement targets are labeled by ID number mddnn; in the K$'$
images target positions are predicted from constant drift models.  The
morphology and apparent position of features can differ in K$'$ and H
images due to their different vertical sensitivities.  The map grid is
at intervals of 30\deg in longitude and 10\deg in planetocentric
latitude. In the 28 July images, the uncircled bright objects below
the equator are the satellites Puck and Miranda.}
\label{Fig:2x3jul27}
\end{figure*}

\begin{figure*}\centering
\includegraphics[width=5in]{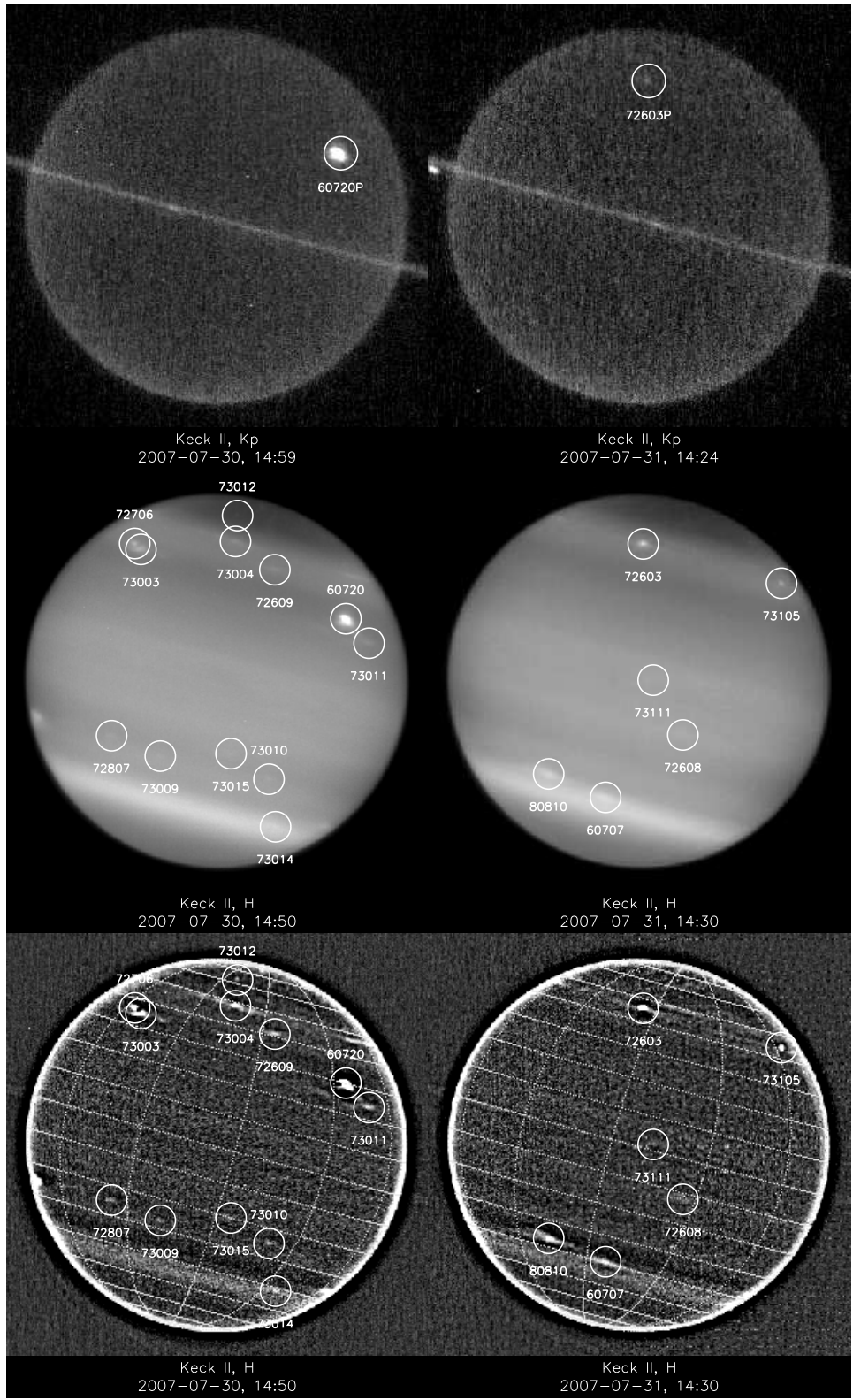}\par
\caption{As in Fig.\ \ref{Fig:2x3jul27} except that images were
made on July 30 and 31.}
\label{Fig:2x3jul30}
\end{figure*}

\begin{figure*}\centering
\includegraphics[width=5in]{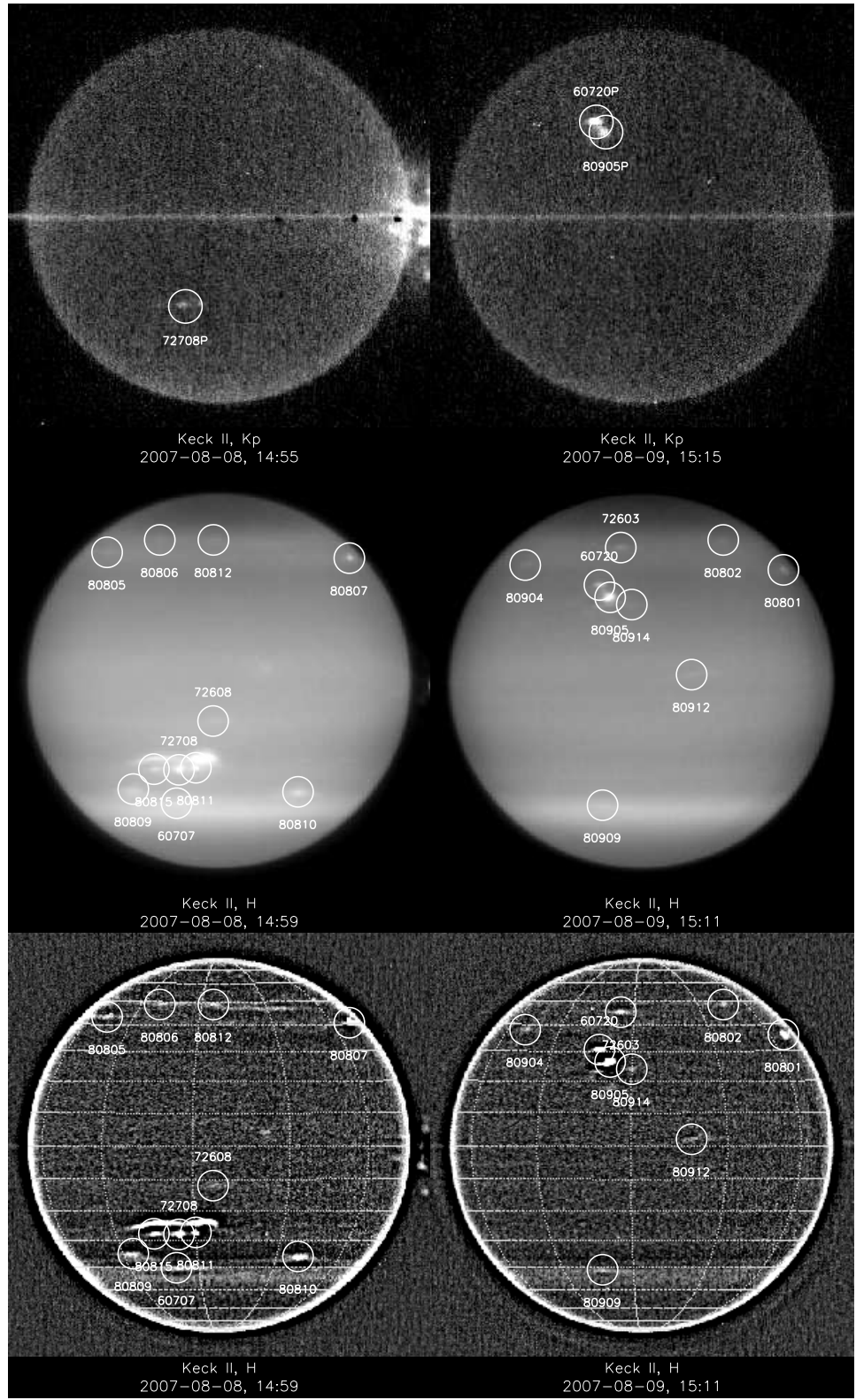}\par
\caption{As in Fig.\ \ref{Fig:2x3jul27} except that images were
made on August 8 and 9.  Note the rare appearance of a southern
cloud feature in the K$'$ image.}
\label{Fig:2x3aug8}
\end{figure*}

\begin{figure*}\centering
\includegraphics[width=5in]{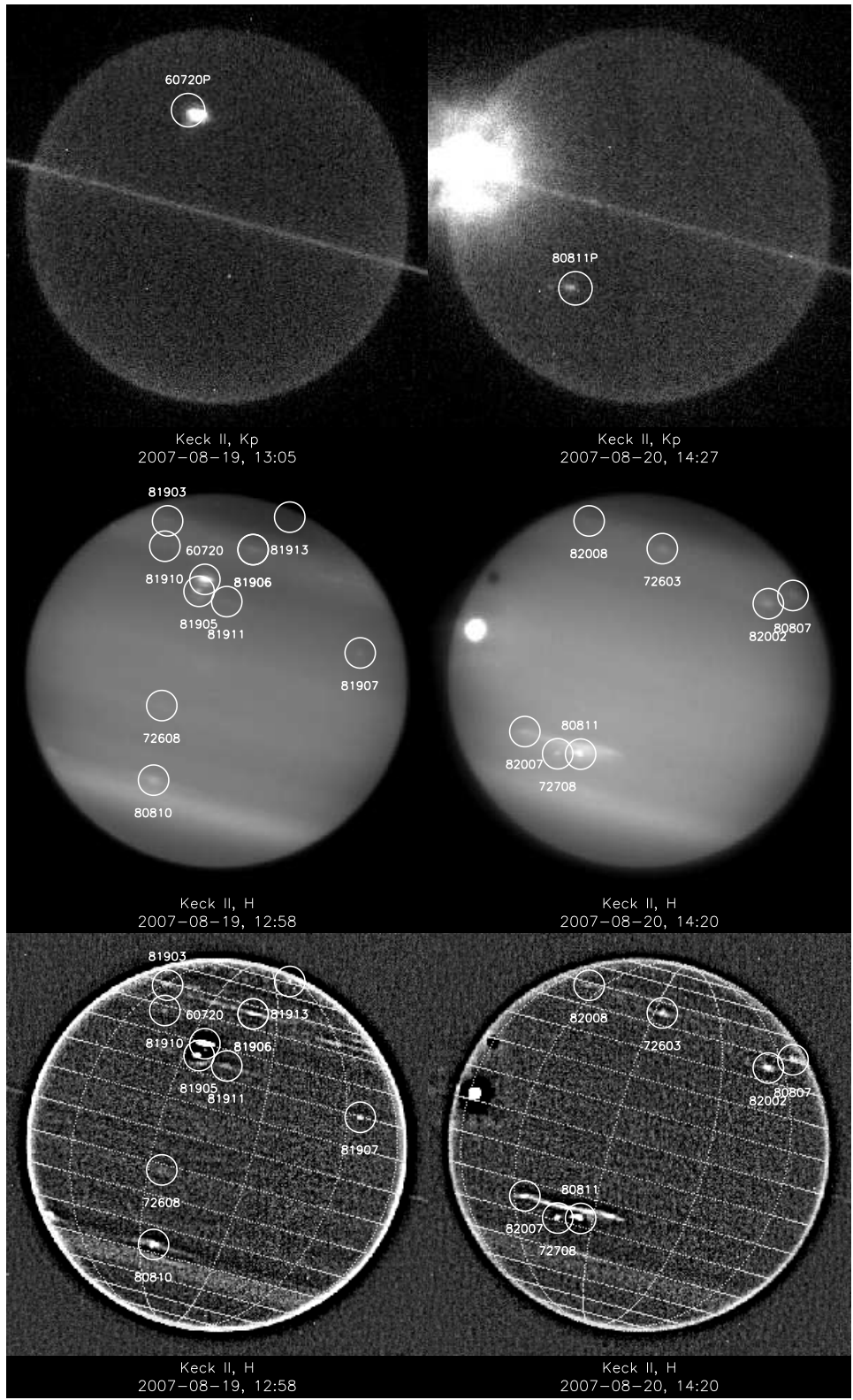}\par
\caption{As in Fig.\ \ref{Fig:2x3jul27} except that images were made
on August 19 and 20.  Note another rare appearance of a southern cloud
feature in the K$'$ image.  The bright object near the left limb in
the 20 August image is the satellite Titania. The dark spot is its shadow.}
\label{Fig:2x3aug19}
\end{figure*}

\begin{figure*}\centering
\includegraphics[width=5in]{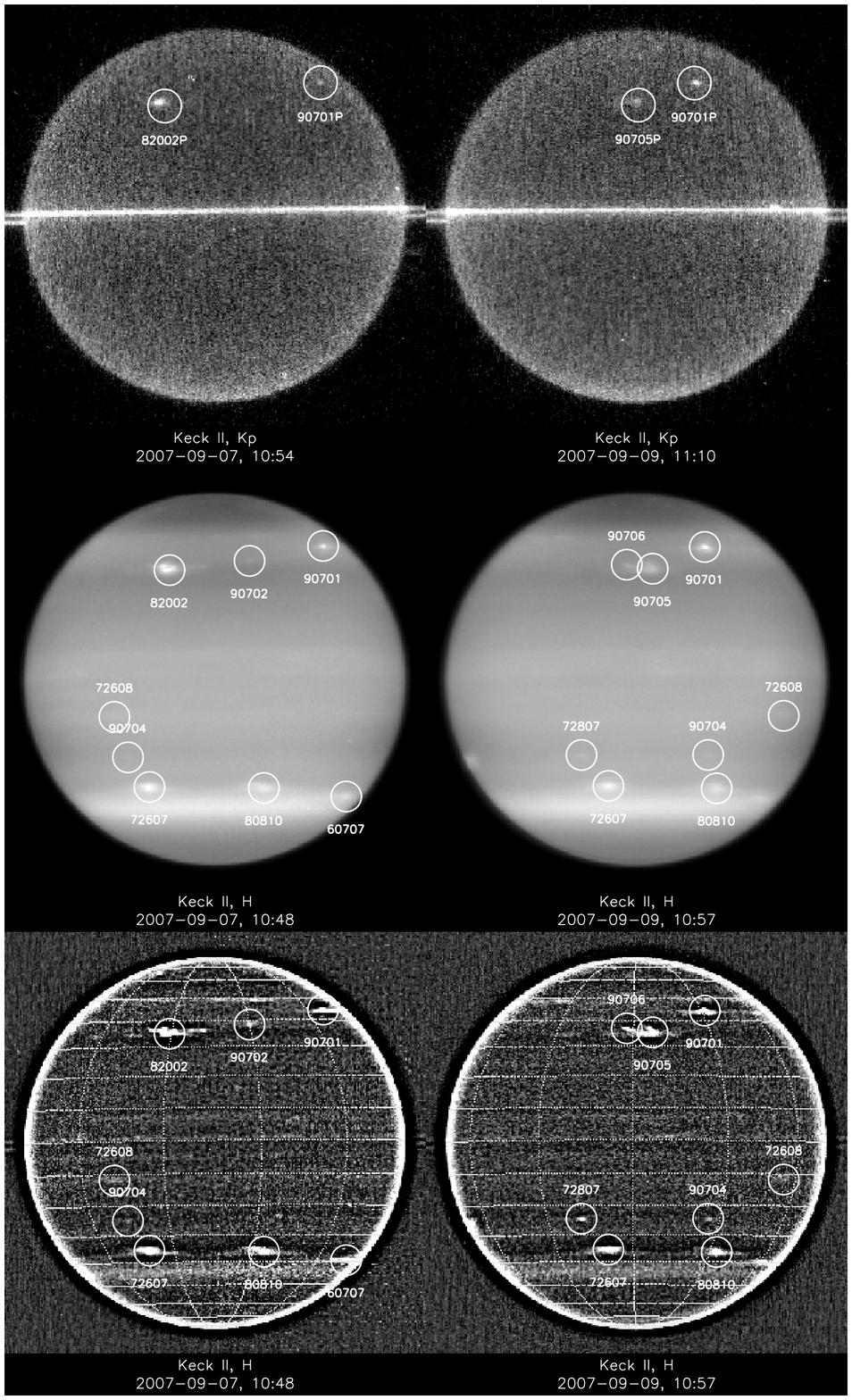}\par
\caption{As in Fig.\ \ref{Fig:2x3jul27} except that images were
made on September 7 and 9. This pair differs from previous pairs
by being taken at a time difference of two days, which is 
roughly three rotations of Uranus, allowing several targets
to be seen on both nights, while the other images, being
separated by roughly 1.5 Uranian rotations showed mainly
opposite hemispheres.}
\label{Fig:2x2sep7}
\end{figure*}

A sampling of the discrete cloud features we used to track atmospheric
motions is provided for all but the June 7 observing run in Figs.\
\ref{Fig:2x3jul27}- \ref{Fig:2x2sep7}, where K$'$ and H images are
shown along with a high-pass filtered H image. (June 7 did not add any
features not seen on other dates.)  All the encircled cloud targets
were observed on multiple images and were used to determine drift
rates and wind speeds. Target labels have the form mddnn, where m and
dd denote the month and day when a given target was prominent and
clearly identified, and nn is the sequence number for the date.
Earlier and/or later observations of the same feature were often
identified and linked together for estimating wind speeds, in which
case all such targets were given the same label, eliminating some of
the original labels given before the linkage was established.  Only
the higher altitude cloud features ($P <1$ bar) are visible in the
K$'$ images, and almost all of those are seen in the northern
hemisphere.  The reason target circles in K$'$ images are sometimes
displaced from target locations is that in these images the positions
are based on predictions of a constant drift model, while many of the
clouds actually move at a variable drift rate.  

Many of the cloud targets were quite subtle, and were most easily
discerned in high-pass filtered images, shown in the bottom rows of
the aforementioned figures. The key methods of distinguishing subtle
cloud features from noise are the time dependence of position and
morphology.  A trackable cloud must not change shape significantly
over the tracked interval and must have a consistent drift rate over
the tracked interval.  The uncertainties in tracking are revealed by
comparison of measured positions relative to our best-fit constant
drift track, which is only meaningful with 3 or more time samples.  As
discussed later, all but one target are sampled 4 or more times, and
many far more than that. Uncircled bright features south of the
equator in Figs.\ \ref{Fig:2x3jul30} and \ref{Fig:2x3aug19} are
satellites.

Keck II imaging observations in the H filter provided the best
combination of contrast and resolution, and thus provided most of the
input to the cloud tracking effort.  Most of our Keck runs provided
observations on two successive nights, between which Uranus undergoes
1.39 rotations.  As a result, there were only a few cloud targets seen
on both nights of each observing run. These provided the highest
initial wind speed accuracy, of the order of 1.3 to 3 m/s. Most
targets were followed during only a single transit on one night. The
best accuracy for these is about 10 m/s, for a target tracked over
about 4 hours.  Typical errors are twice this large because most
clouds are tracked for much less than four hours.  The longest
tracking time within a single transit was 4.17 hours.  The estimated
uncertainties in the zonal winds are of the same order as
uncertainties in the meridional winds, although meridional wind speeds
are much smaller.  Because we could not resolve any systematic
meridional motion for features observed on only one transit,
meridional wind speeds are not tabulated.  However, rather substantial
meridional excursions of $\pm$2.1\deg of latitude were observed in
prior years for a long-lived feature near 34\degx S.  Empirical fits
to a simple inertial oscillation model imply that this feature
regularly reaches peak meridional speeds of 28 m/s \citep{Sro2005dyn},
but the speed varies so rapidly that it is difficult to resolve by
direct measurement.

In many cases we were able to find cloud features that lasted long enough to
track over very extended time periods, the longest of which was 2255 hours
(94 days).  We tracked 15 targets for more than 100 hours, seven of which
were tracked for more than 1000 hours.  Many of these moved at extremely
uniform drift rates, with an average speed defined to within a few cm/sec,
and average drift rates to within 10$^{-4}$\degx /h. 

\subsection{Cloud motion fits and uncertainties}

Our wind results are obtained by fitting individual longitude
observations to a constant drift model.  Two fits are computed: one
weighting individual observations by the inverse of their expected
variances, and the second is an unweighted fit. In the latter case the
standard deviation of the points from the fit provides the basis for
the error estimate and accounts for uncertainties in identification
due to inherently poor morphology (long streaky features) or
variations in contrast or in feature evolution.  The weighted error
estimates are computed in the pixel domain, with individual position
errors taken to be 0.6 pixel RMS, an estimate that roughly accounts
for the observed variability in measurements over a large sample, and
thus incorporates both typical tracking and image navigation
errors. Our final wind result is the average of the two fits, and
assigned the larger of the two errors.  Individual measurement
uncertainties are typically 0.3\deg of latitude and longitude near the
central disk and increase towards the limbs.

Over a single transit, fitting errors are generally smaller than our
{\it a priori} estimates ($\chi^2 < N-2$) and over longer periods the
errors are close to our {\it a priori} values for latitude, but about
twice as large for longitudinal measurements.  The increased errors in
longitude are partly due to the longitudinally stretched character of
many features, making them difficult to locate in that dimension, and
also because systematic deviations in drift rate can accumulate much
larger longitudinal deviations than the corresponding systematic
variations in latitude.  Thus, part of what seems like longitudinal
measurement error is really just systematic deviation from our simple
constant drift model. Some cloud feature motions exhibit short period
and long period oscillations, leading to systematic deviations in both
longitude and latitude from constant longitudinal drift at a fixed
latitude. Only a few cloud targets are sampled well enough to
characterize these systematic variations.

The uncertainties quoted for both mean positions and mean drift rates
are formal uncertainties based on an assumed random distribution of
errors, with no accounting for possible systematic deviations from the
assumed constant drift models.  For large numbers of measurements over
long time intervals, these formal uncertainties can become quite small
and do not necessarily (more often do not) accurately reflect the
steadiness of the motion or position of the tracked feature, nor our
ability to predict where it will be in the future.  Besides unsteady
motions, there are two other effects that might produce significant
systematic errors in latitude: pixel scale errors and center-finding
errors associated with north-south asymmetries in limb profiles,
either because of physical differences in the atmosphere or
differences in illumination at non-zero phase angles. By measuring
satellite positions we were able to verify the published NIRC2 pixel
scale to better than 0.4\%. An error of this size would not affect our
center finding accuracy, but would cause a latitude error that
increased towards the north and south limbs, reaching values of
0.13\degx, 0.38\degx, and 0.61\degx at latitudes of 30\degx, 60\degx,
and 70\degx respectively.  To assess the limb-asymmetry impact, the
limbs of images at 1.0\deg and 2.9\deg phase angles were fit to model
ellipses first using all available points, then fitting only the left
or right (south or north) limb. 
The largest difference between fitting both limbs together, which is
our standard method, and fitting just one limb was only 0.34 pixels,
which is a reasonable bound on the systematic center-finding error in
the latitude direction.  At latitudes of 0\degx, 30\degx, 60\degx, and
70\degx, the corresponding errors in latitude would be 0.11\degx, 0.13\degx,
0.22\deg, and 0.32\deg respectively.

\subsection{Cloud-tracked wind results}

 We follow \cite{Allison1991uranbook}, \cite{Hammel2001Icar,
Hammel2005winds}, and \cite{Sro2005dyn} in using prograde (IAU
westward) winds as positive winds on Uranus. On most planets, eastward
winds would be prograde.  Uranus is unusual because the
rotational pole of Uranus (as defined by the right hand rule) is 98\deg from
its orbital pole. Because this puts the rotational pole south of the
invariable plane of the solar system, IAU convention makes this the
south pole of the planet \citep{Seidelmann2002}.

\subsubsection{Tracking discrete features}

From observations during a single transit we obtained 51 cloud-tracked
winds with wind speed uncertainties less than 40 m/s, the majority of which
being less than 20 m/s. A table of these results can be obtained
as on-line supplementary material at www.ssec.wisc.edu/planetary/uranus/onlinedata/
ura2009eqdyn/. In Table\ \ref{Tbl:ulong}, we summarize
the high-accuracy results of tracking features over more than a single transit.
Remarkably, we were able to find 28 such features, thanks to the
relatively close spacing of observations during the period approaching
equinox.  We found these long-lived targets by linking together
observations over shorter time intervals, using preliminary fits from
short time intervals to identify possible cases in which two targets
were actually the same cloud feature.  After linking two such
measurement groups together, the longer time base provided much more
accurate predictions that facilitated linkages over longer time
intervals.  Besides the consistency of longitudinal tracks, latitude
and morphology were also used as a basis for identification of
features.  In some cases two features appeared close to the predicted
latitude and longitude of another feature, in which case we did not
attempt a linkage.

\begin{figure}\centering
\hspace{-0.34in}\includegraphics[width=3.73in]{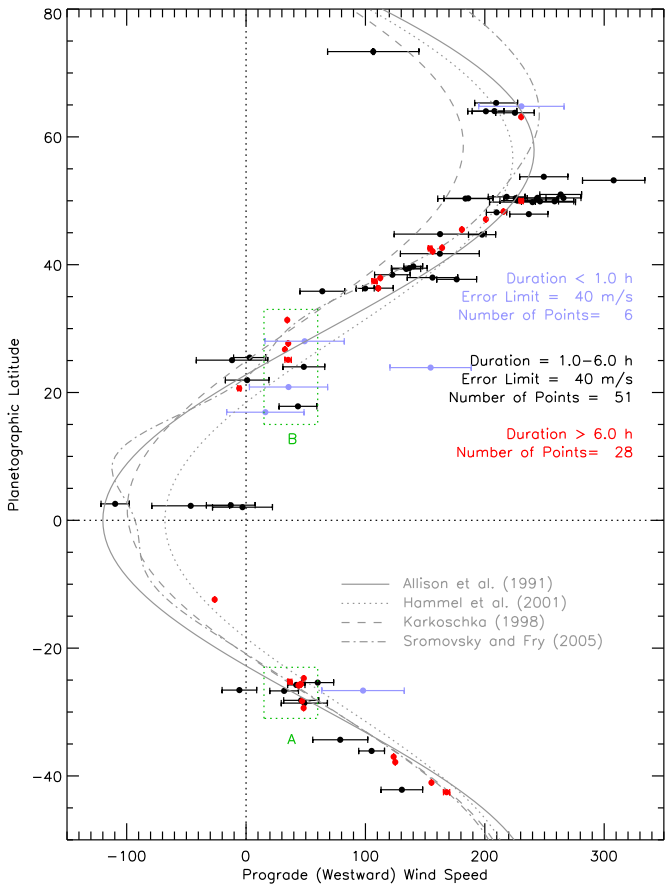}\par
\caption{Zonal wind results vs. planetographic latitude, showing three subsets:
short-duration ($<$ 1 h, light blue) observations, single-transit (1-6 h, black), and
long-duration ($>$6 H, red) observations (red). Model curves (grey) are discussed in
the text. Regions A and B contain observations from major circulation features.}
\label{Fig:ugroups}
\end{figure}

All the wind determinations yielding zonal speed uncertainties $\le$40
m/s are plotted versus latitude in Fig.\ \ref{Fig:ugroups}.  Six of
these, measured over time spans less than 1 hour, are shown as light
blue. The 4$\sigma$ deviant measurement of this subset, near 20\deg N and
150 m/s was tracked over an interval of less than 1 hour, with all but
one of 14 measurements within 24 minutes.  It is just south of what
appears to be a dark vortex and is probably a companion to that
feature.  Evolution of companions is common and likely contributed to
tracking errors that resulted in this substantial deviation. The more
reliable wind values from single transit tracking in the 1-6 hour
range are shown in black.  The highly accurate long duration tracking
results are shown in red.  Although error bars are plotted for all
results, the long-duration error bars are generally too small to
notice.

\begin{table*}\centering
\caption{Zonal wind fits for 28 clouds tracked for more than 19 hours.}
\vspace{.125in}
\renewcommand{\baselinestretch}{1.0}
\small
\begin{tabular}{r r r r c c r}
   &     & Eastward & 
zonal wind & ID &  & track\\

Lat. ($\phi$) & Lat. ($\phi_{pg}$) & drift rate (\degx /h) & 
 (m/s west) & (mddnn) & N$_\mathrm{obs}$ & time (h)\\
\hline
   62.03&   63.12$\pm$0.10&  -4.0383$\pm$0.0041&  230.45$\pm$0.25&  72810&   18&    74.74\\
   48.66&   49.97$\pm$0.13&  -2.8492$\pm$0.0144&  230.56$\pm$1.76&  80802&    5&    27.95\\
   46.99&   48.31$\pm$0.06&  -2.5844$\pm$0.0020&  215.64$\pm$0.17&  72601&   22&   123.34\\
   45.78&   47.11$\pm$0.04&  -2.3531$\pm$0.0002&  200.90$\pm$0.07&  72603&   47&  1082.70\\
   44.17&   45.50$\pm$0.08&  -2.0563$\pm$0.0042&  180.83$\pm$0.37&  90701&   13&    50.56\\
   41.31&   42.64$\pm$0.06&  -1.7826$\pm$0.0007&  164.20$\pm$0.17&  72609&   18&   334.94\\
   41.22&   42.54$\pm$0.08&  -1.6621$\pm$0.0049&  154.19$\pm$1.50&  72803&   10&    68.23\\
   40.76&   42.08$\pm$0.07&  -1.6824$\pm$0.0010&  156.27$\pm$0.09&  80807&   13&   287.62\\
   36.64&   37.92$\pm$0.05&  -1.1441$\pm$0.0033&  112.40$\pm$0.34&  72706&   20&    72.12\\
   36.14&   37.42$\pm$0.10&  -1.0888$\pm$0.0211&  107.52$\pm$2.09&  80801&    5&    27.95\\
   35.05&   36.31$\pm$0.07&  -1.1005$\pm$0.0005&  110.77$\pm$0.05&  82002&   10&   456.25\\
   30.17&   31.34$\pm$0.03&  -0.3240$\pm$0.0001&   34.50$\pm$0.04&  60720&   63&  1751.44\\
   26.60&   27.68$\pm$0.05&  -0.3202$\pm$0.0004&   35.30$\pm$0.10&  72606&   21&   340.57\\
   25.68&   26.73$\pm$0.07&  -0.2926$\pm$0.0005&   32.53$\pm$0.06&  81905&    8&   458.81\\
   24.10&   25.11$\pm$0.07&  -0.3097$\pm$0.0100&   35.25$\pm$2.69&  73011&    8&    19.87\\
   19.79&   20.65$\pm$0.08&   0.0504$\pm$0.0112&   -5.75$\pm$1.30&  81907&    6&    21.64\\
  -11.86&  -12.40$\pm$0.03&   0.2171$\pm$0.0001&  -26.31$\pm$0.02&  72608&   39&  1082.66\\
  -23.72&  -24.71$\pm$0.06&  -0.4274$\pm$0.0003&   48.38$\pm$0.27&  72712&   14&   577.16\\
  -24.25&  -25.26$\pm$0.12&  -0.3262$\pm$0.0148&   36.73$\pm$1.67&  72714&    3&    19.62\\
  -24.61&  -25.63$\pm$0.04&  -0.4094$\pm$0.0001&   45.95$\pm$0.03&  72807&   30&  1080.41\\
  -24.73&  -25.75$\pm$0.09&  -0.4038$\pm$0.0044&   45.25$\pm$0.59&  90704&    5&    50.50\\
  -24.77&  -25.80$\pm$0.05&  -0.3923$\pm$0.0016&   43.90$\pm$0.62&  73010&   17&   100.83\\
  -27.17&  -28.27$\pm$0.05&  -0.4287$\pm$0.0003&   46.99$\pm$0.08&  80811&   15&   764.12\\
  -28.25&  -29.37$\pm$0.05&  -0.4442$\pm$0.0002&   48.22$\pm$0.04&  72708&   23&   577.20\\
  -35.73&  -37.00$\pm$0.04&  -1.2388$\pm$0.0001&  123.63$\pm$0.02&  72607&   29&  1078.67\\
  -36.56&  -37.84$\pm$0.03&  -1.2661$\pm$0.0001&  124.97$\pm$0.05&  80810&   50&  1055.93\\
  -39.74&  -41.05$\pm$0.03&  -1.6459$\pm$0.0001&  155.32$\pm$0.05&  60707&   54&  2254.79\\
  -41.22&  -42.54$\pm$0.09&  -1.8206$\pm$0.0188&  167.98$\pm$2.62&  73014&    8&    19.87\\
\hline
\end{tabular}\label{Tbl:ulong}
\parbox[]{4.75in}{NOTE: Uncertainty estimates here exclude the systematic errors discussed in Sec. 3.2.}
\renewcommand{\baselinestretch}{2.0}
\normalsize
\end{table*}

In Fig.\ \ref{Fig:ugroups}, four model zonal wind profiles are
provided for reference.  The solid curve is a symmetric fit to the
1986 Voyager wind observations by \cite{Allison1991uranbook}.  This
weights all Voyager observations equally, and thus probably gives too
much weight to the highly uncertain near-equatorial determination
based on radio occultation results of \cite{Lindal1987}.  The second
symmetric curve (dotted) by \cite{Hammel2001Icar} provides a better
fit to the more accurate cloud-tracked winds obtained by Voyager
imaging \citep{SmithBA1986}. Equations for these two fits are given by
\cite{Hammel2001Icar}. The third curve (dashed) is an asymmetric fit
by \cite{Kark1998Sci} based on HST observations that provided
measurements in northern latitudes not visible to Voyager cameras. 
The fourth curve (dot-dash) is an asymmetric fit by
\cite{Sro2005dyn} to a combination of 2003-4 Keck observations and
Voyager and Karkoschka HST results.

Our new observations extend to higher northern latitudes than prior
observations (up to 73\deg N), which allows us for the first time to
define the location of a northern hemisphere prograde jet peak
at roughly 58\deg N, which is now much better defined than
the southern jet and crudely symmetric to it.
Our new results also provide a large number of highly accurate
wind measurements, which indicate that the current wind profile is
asymmetric, roughly following the \cite{Sro2005dyn} profile, except at
low latitudes and in two special regions outlined in green boxes.  In
both outlined regions the highly accurate determinations (red)
indicate regions of nearly zero latitudinal gradients in wind
speed.  The most plausible explanation for this is that several of the
tracked features are generated by a single circulation feature, much
as the bright companion clouds generated by the Great Dark Spot on
Neptune \citep{Sro2002spots}.  These bright companions (on Neptune) are thought to
be orographic clouds produced by vertical motions as the zonal flow is
deflected around the vortex that creates the dark spot
\citep{Stratman2001}.  Cloud groupings are further discussed in the following
section. 

\subsubsection{Cloud complexes}

\begin{figure*}\centering
\includegraphics[width=6in]{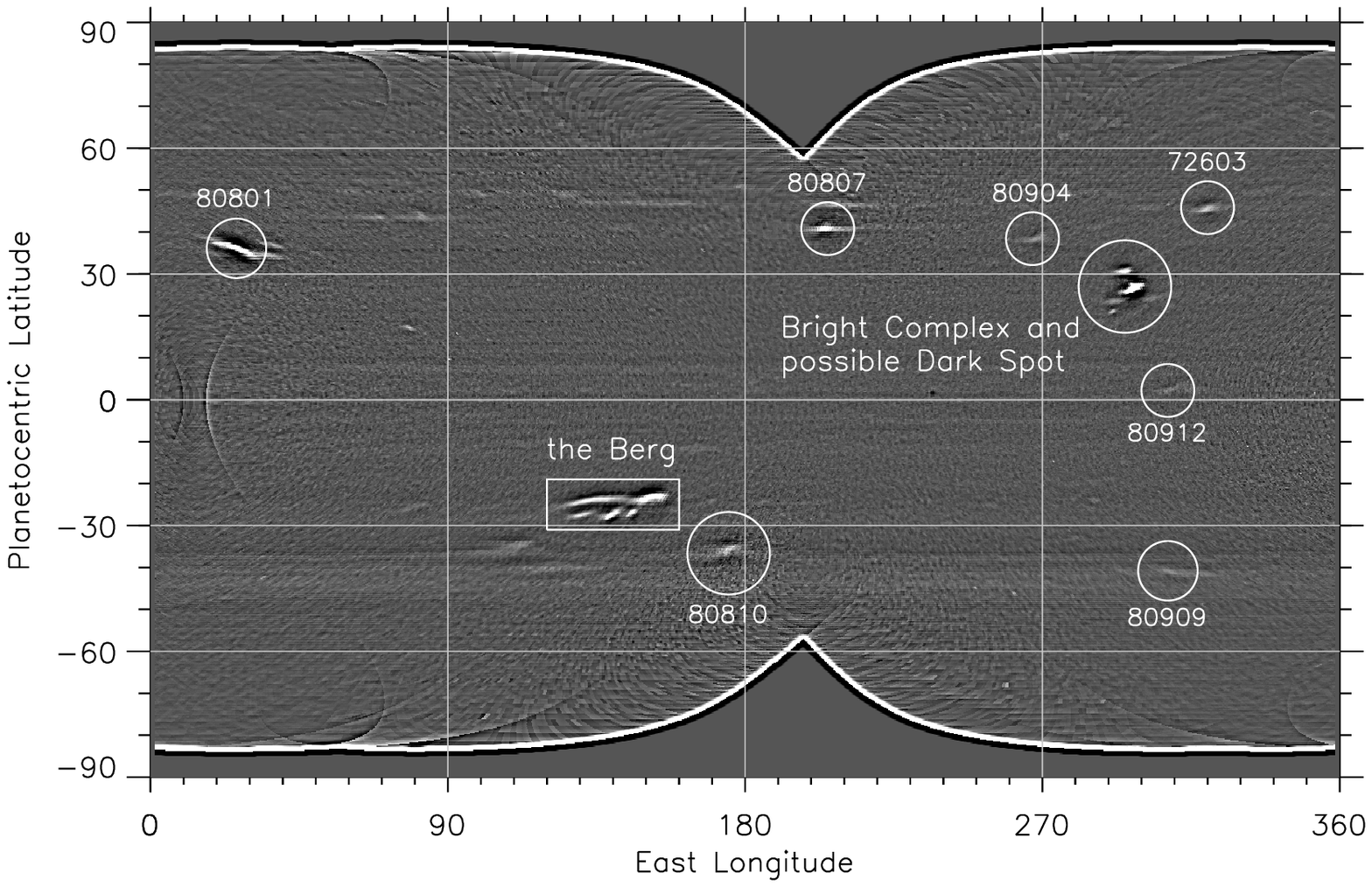}\par
\caption{Mosaic of rectilinear projections of Uranus images made on 8
and 9 August 2007 by the Keck II NIRC2 camera using an H filter,
displayed here with a high-pass filter to enhance the contrast of
subtle cloud features.  The two main collections of bright features
with components that exhibit group longitudinal motions are the Berg
and the Bright Complex, the latter likely associated with the dark
spot seen best at visible wavelength.}
\label{Fig:mosaic}
\end{figure*}

The groups of bright cloud features associated with the two outlined
regions in Fig.\ \ref{Fig:ugroups} are put into context by the
mosaicked rectilinear projection of images taken on 8-9 August 2007
(Fig.\ \ref{Fig:mosaic}), in which it can be seen that nearly all of
cloud features in the outlined latitude bands are also grouped
together in close spatial proximity.  These spatial groups are labeled
Berg and Bright Complex for reasons discussed below.  In the latitude
regions occupied by the closely grouped features, there are very few
accurate wind vectors that can be obtained at other longitudes that
might be used to define the true latitudinal gradient of the zonal
wind in these latitude bands.

In 2006, Hubble Space Telescope images revealed a small dark spot in
the atmosphere of Uranus \citep{Hammel2008spot}, which was found to
have associated bright companions.  The dark feature was tracked in
HST images and the companion groups visible at near-IR wavelengths
were tracked in both HST and groundbased images.  In fact, the feature
generating cloud targets in region B (Fig.\ \ref{Fig:ugroups}) is
possibly the same feature identified as a dark spot at visible
wavelengths in 2006, and may be the same feature that produced the
brightest cloud feature ever observed on Uranus in 2005
\citep{Sro2007bright}, termed the Bright Complex. It is also
conceivably feature V reported by \cite{Hammel2005winds} and also the
unusually bright feature identified by \cite{Sro2000}.  The latitude
of the presumed vortex feature in the 2007 observations is likely
close to the middle of the group of red points in region B.  This is
somewhat uncertain because the presumed vortex feature does not have
well-defined boundaries and may undergo latitude excursions.
Companions can appear in different locations relative to a vortex and
also can vary with time.  Thus the rough center of a group of
companions is only an estimate of where the vortex might be located.

As can be surmised from Figs.\ \ref{Fig:2x3jul30}-\ref{Fig:2x2sep7},
trackable discrete cloud features tend to cluster in narrow latitude
bands, leaving several gaps of $\sim$10\deg at low latitudes, and
larger gaps at high latitudes.  In northern high latitude regions,
which are just beginning to be seen, there appear to be small discrete
cloud features that should provide very usable targets in coming years
as the observer subpoint moves further northward. On the other hand,
from 44\deg S (planetocentric) to the south pole, we have not seen a
discrete cloud feature since 1986. The only discrete feature ever
tracked in this region was at 71.7\deg S moving at 175$\pm$4 m/s and
seen only in Voyager 2 UV images \citep{Smith1986}.

\begin{figure*}\centering
\includegraphics[width=6in]{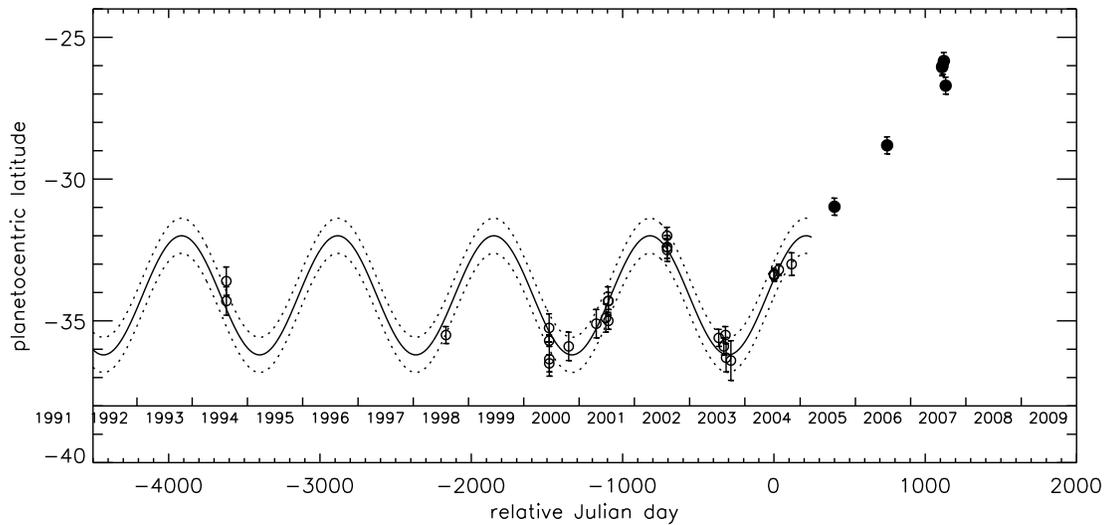}\par
\caption{Long term motion of the Berg, previously confined to
32\degx S - 36\degx S (1994-2004 points from \citep{Sro2005dyn}, but
recently moving substantially northward (filled circles) and
undergoing large variations in morphology (Fig.\ \ref{Fig:berg}). The
latitudes from 2005 forward are for the small bright spot, and are
about 1\deg S of the central latitude of the Berg complex, which is
what was plotted for prior years.}
\label{Fig:berglat}
\end{figure*}

The wind results in region A of Fig.\ \ref{Fig:ugroups} are derived
from clouds that are all associated with a major circulation feature
that may have been present since 1986. The 2007 group of targets
includes 60720, 72603, 72606, 72817, and 73011, which are identified
in Figs.\ \ref{Fig:2x3jul27}, \ref{Fig:2x3jul30}, and
\ref{Fig:2x3aug8}.  The long-term behavior of this cloud complex was
first described by \cite{Sro2005dyn}, who referred to it as S34
because it had for many years been oscillating between planetocentric
latitudes of 32\deg S and 36.5\deg S, possibly for two decades.  But
since it began drifting equatorward in 2005 (Fig.\ \ref{Fig:berglat})
the informal name Berg seems more appropriate (this arose from its
vague similarity to an iceberg disconnected from an ice shelf).

\subsubsection{Bin-averaged single-transit winds}

To improve wind speed accuracy in latitude bands with a high density
of single-transit cloud targets, we binned the 1-6 hour tracking
results into 2\deg latitude bins and computed weighted averages. The
results for 2\deg bins, including only bins for which estimated errors
could be reduced to 10 m/s or better are provided in Table\
\ref{Tbl:ubin}. For these 10 bins, the listed latitude is the mean
latitude of the wind vectors included in the bin, not the latitude of
the center of the bin.  The binned and high-accuracy tracking results
are plotted in Fig.\ \ref{Fig:legfit}, where we also show two
different fits to the observations, which are described in the
following section.

\subsubsection{Fitting smooth functions to zonal wind observations}

To facilitate use of these results by the scientific community, we fit
the raw observations with 10th order and 11th order Legendre
polynomial expansions. These polynomials are orthogonal solutions to
Laplace's equation in spherical coordinates, ideally suited to
representation of a zonally uniform function, and are used by
dynamical modelers \citep{Lebeau1998}. Because of large latitude gaps
in the observations, fits of these and higher orders are prone to wild
oscillations in the unmeasured regions. Also, because of the extremely
high accuracy of some of the long-term observations, a fit that
strictly weights observations according to estimated measurement
accuracy can be dominated by a few highly accurate observations at the
expense of large deviations in regions with few accurate measurements.
To provide a degree of smoothing in the fitting process and to
equalize weighting over all measured latitudes, we applied the
following constraints: each 2007 observation had its error increased
by root-sum-squaring it with a minimum error of 1 m/s; in addition,
synthetic measurements of 0.0 m/s$\pm$0.01 m/s were added at the
poles, and the Voyager result at 71.7\deg S (175$\pm$4 m/s) was added
to inhibit instability in what would otherwise be a large unmeasured
region and because it is the only measurement ever obtained between
the south pole and 42\deg S.  The resulting fits are shown by the
solid curves in Fig.\ \ref{Fig:legfit}, with fit coefficients given in
Table\ \ref{Tbl:fits}. The relative $\chi^2$ values of these fits are
1440 and 1280 for the order 10 and 11 Legendre fits respectively.
These large values of $\chi^2$ are a result of the regions of nearly
zero wind speed gradients, which cannot be well fit with the imposed
smoothing constraints.  The dispersion of these different fits in
regions without measurements is a useful reminder that the true
profile there is not constrained by observations.

\subsubsection{North-south asymmetry}

\begin{figure}\centering
\hspace{-0.2in}\includegraphics[width=3.55in]{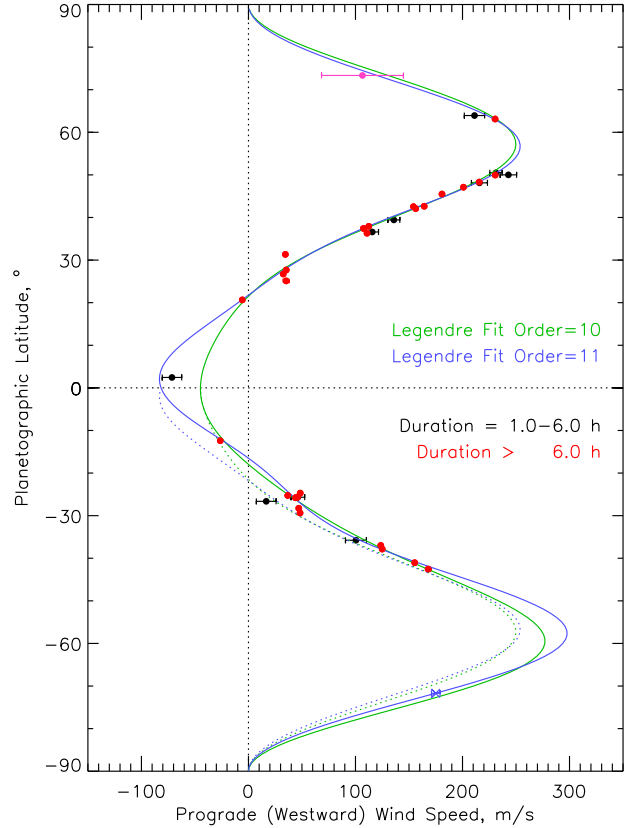}
\caption{Wind measurements obtained from 2007 Keck II imaging
observations. Long duration results (28 red points) and binned 1-4.2 h
results (10 black points) are fit with 10th and 11th order Legendre
polynomial series (green and blue curves). Binned averages used 51
points in 2\deg bins. Fit coefficients are given in Table\
\ref{Tbl:fits}.  Northern fits are also plotted with latitudes
inverted (dotted curves) to show profile asymmetries with respect to
latitude. The pink point at 73.4\deg N, not used in the fits because
of its uncertainty, is our highest latitude observation.  The blue
point at 71.7\deg S is due to \cite{SmithBA1986}.}
\label{Fig:legfit}
\end{figure}

\begin{figure}\centering
\hspace{-0.2in}\includegraphics[width=3.55in]{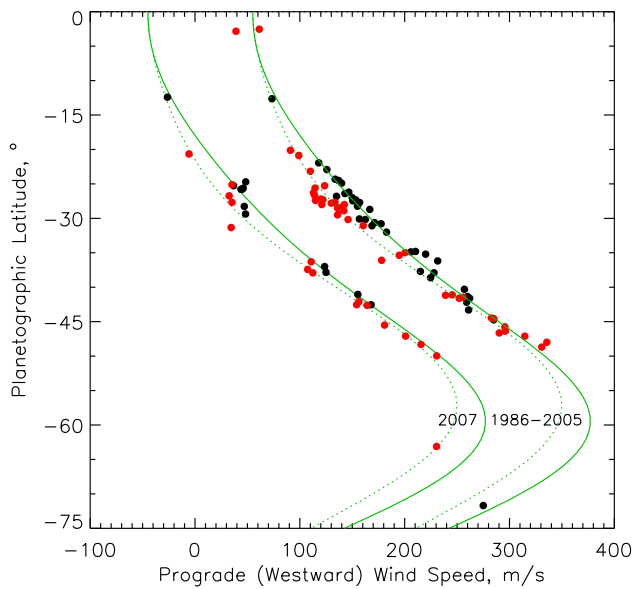}
\caption{Comparison of high-accuracy wind observations (black dots)
with northern observations reflected about the equator (red dots). The
left group is from 2007 (this work) and the right group (displaced 100
m/s to the right) combines results from 1986 through 2005.  For
reference, we show the 10$^\mathrm{th}$ order Legendre fit in green,
both displaced and undisplaced, with dotted curves showing reflections
of the northern fit. Here we include only measurements with wind
errors $<$10 m/s and latitude errors $<$ 0.5\degx.}
\label{Fig:asymm}
\end{figure}

The fit profiles compared to their reflections about the equator
(dotted curves in Fig.\ \ref{Fig:legfit}) indicate that the wind
profile is asymmetric. The biggest asymmetry may be in the region of
the prograde jets near latitudes of $\pm$54\degx, but at present this
is hardly constrained at all by the observations.  Although the
northern jet will likely become better defined within the next several
years, presuming that an obscuring south polar cap does not form too
rapidly, there is less prospect for improved measurements in the
southern band region, or especially poleward, before the next equinox
(in 2049), although there is some hope that the southern collar may
break up into trackable features before disappearing entirely.  The
asymmetry is more convincingly shown by direct comparison of
observations. In Fig.\ \ref{Fig:asymm}, we plot the most accurate
observations from 2007 in black, then plot the northern half of the
observations with inverted latitudes in red, providing a direct
comparison of northern and southern observations at the same distance
from the equator.  We also show a similar plot for all observations
from 1986 to 2005, displaced by 100 m/s to avoid overlap.  This second
curve is dominated by the more recent measurements from 1997-2005,
which will be further discussed in the following section.  Although the
asymmetry is not large, it is clearly resolvable and consistently
supported by recent measurements.  The best defined asymmetry is in
the 20-40\deg latitude range, where it is only about 20 m/s.

\begin{table*}\centering
\caption{Binned zonal winds for 1-4.2 hour cloud targets, 2\deg bins, and uncertainties $\le$ 10 m/s.}
\begin{tabular}{r r c c r c r}
\\ & & Std. Dev.& Eastward & Zonal wind&  \\
  Lat. ($\phi$) & Lat. ($\phi_{pg}$) & in Lat. (\degx) & drift
              rate (\degx /h) & (m/s westward) & NBIN\\
              \hline
   62.87&   63.93$\pm$0.18&  0.12& -3.80$\pm$0.17&  211$\pm$10&     3\\
   49.19&   50.50$\pm$0.14&  0.19& -2.84$\pm$0.07&  231$\pm$06&      7\\
   48.70&   50.01$\pm$0.13&  0.15& -3.01$\pm$0.10&  243$\pm$08&      5\\
   46.80&   48.13$\pm$0.08&  0.11& -2.58$\pm$0.09&  216$\pm$08&     2\\
   38.13&   39.43$\pm$0.10&  0.43& -1.41$\pm$0.06&  136$\pm$06&     4\\
   35.34&   36.60$\pm$0.08&  0.60& -1.25$\pm$0.05&  116$\pm$06&     4\\
    2.33&    2.44$\pm$0.09&  0.18&  0.45$\pm$0.06&  -71$\pm$09&     4\\
  -24.65&  -25.67$\pm$0.08&  0.15& -0.42$\pm$0.06&   46$\pm$06&      2\\
  -25.59&  -26.64$\pm$0.09&  0.06& -0.15$\pm$0.08&   17$\pm$09&     2\\
  -34.52&  -35.78$\pm$0.09&  0.66& -0.96$\pm$0.07&  100$\pm$10&     2\\
\hline
\end{tabular}\label{Tbl:ubin}
\parbox[]{4.75in}{NOTE: Uncertainty estimates here exclude the systematic errors discussed in Sec. 3.2.}
\end{table*}

\begin{table}\centering
\caption{Fits to 2007 measurements of the zonal wind profile of Uranus.}
\renewcommand{\baselinestretch}{1.5}
\small
\vspace{0.15in}
\begin{tabular}{c  c  c}
Legendre        &  Order-10  &   Order-11   \\
Function        &  Coeff.  &   Coeff.    \\
\hline
$P_0(\sin\phi_\mathrm{pg})$ & 79.90  &  75.06   \\
$P_1(\sin\phi_\mathrm{pg})$ &-13.99  & -20.05   \\
$P_2(\sin\phi_\mathrm{pg})$ & 203.5  &  220.4   \\
$P_3(\sin\phi_\mathrm{pg})$ &-159.5  &  4.77    \\
$P_4(\sin\phi_\mathrm{pg})$ &-126.5  & -153.7   \\
$P_5(\sin\phi_\mathrm{pg})$ & -5.24  & -5.307   \\
$P_6(\sin\phi_\mathrm{pg})$ &-11.49  & -110.2   \\
$P_7(\sin\phi_\mathrm{pg})$ & 4.437  &  21.07   \\
$P_8(\sin\phi_\mathrm{pg})$ &-42.05  & -59.91   \\
$P_9(\sin\phi_\mathrm{pg})$ & 14.81  & -47.44   \\
$P_{10}(\sin\phi_\mathrm{pg})$ &     &   28.33  \\
\hline
\end{tabular}\label{Tbl:fits}\par
\vspace{0.1in}
\renewcommand{\baselinestretch}{2.0} \normalsize
\end{table}

\subsection{Stability of the zonal circulation}

\begin{figure}\centering
\hspace{-0.2in}\includegraphics[width=3.55in]{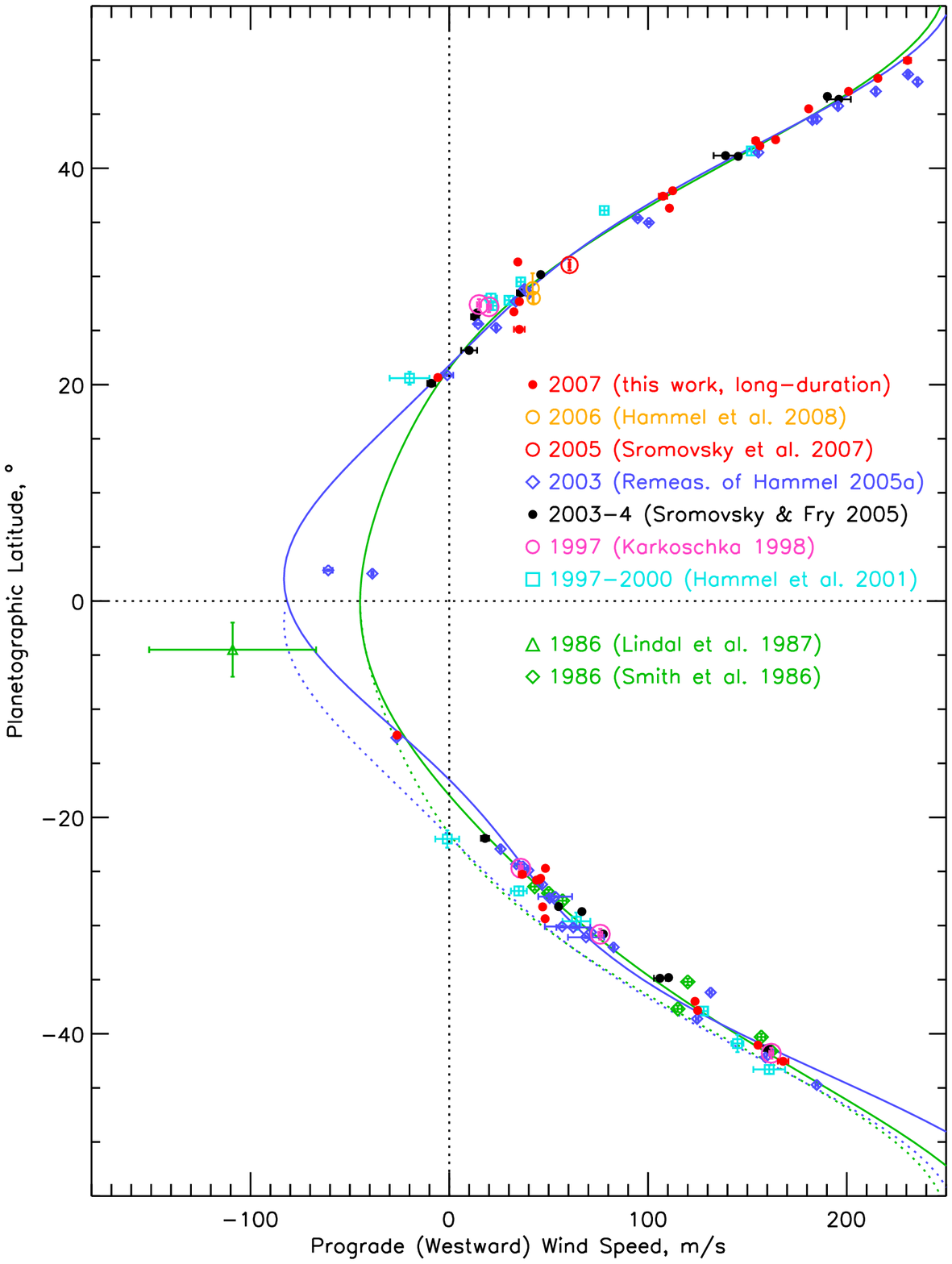}
\caption{Comparison of our 2007-derived Uranus wind profile fits (solid
curves) with current and prior observations given in the legend. The
dotted curves display the reversed northern fits (color coding of
fits as in Fig.\ \ref{Fig:legfit}). See text for discussion.}
\label{Fig:ucomp}
\end{figure}

In Fig.\ \ref{Fig:ucomp} we compare our fit profiles to prior wind
observations of \cite{Smith1986} using Voyager images, radio
occultation results of \cite{Lindal1987}, 1997 HST results by
\cite{Kark1998Icar}, and results by \cite{Hammel2001Icar} that were
derived from a combination of 1998 HST NICMOS images and 2000 HST
WFPC2 and Keck images. We also include high-accuracy and binned
results of \cite{Sro2005dyn}, shown as small black dots, and new
results based on 2003 observations by \cite{Hammel2005winds}, shown as purple
diamonds. We remeasured these latter observations to correct
for a small navigation error that shifted the latitude scale by
$\sim$1.5\degx.  The solid curves shown in this figure are the same
fits given in Fig.\ \ref{Fig:legfit}. Most of the data displayed here
are reasonably consistent with the zonal profile determined from our
2007 observations.  Besides a few deviant points near 27-30\deg N,
which are plausibly related to companion cloud features, the most
significantly deviant points are those of \cite{Hammel2001Icar}.  At
latitudes between 25\deg S and 40\deg N, the \cite{Hammel2001Icar}
observations are $\sim$10 m/s less westward than the trend followed by
most of the other observations, while they are in much closer
agreement elsewhere. 

There is remarkably little difference between the 1986 Voyager results
and the more recent extensive observations in 2003-4 and in 2007.
Perhaps the close agreement of 2003-4 and 2007 results should not be
too surprising because of the relatively short time interval in
comparison to Uranus' rather long radiative time constants
\citep{Conrath1990}. It is much more surprising that there is such
close agreement with the Voyager results, which were obtained 21 years
prior to the 2007 equinox, and close to the midpoint between the last
two equinoxes.  If the asymmetry we see at the time of the 2007
equinox is at a peak, which might be expected if it is a long-delayed
response to seasonal solar forcing, then the Voyager results might be
expected to be half-way between the current asymmetry and its reversal
(shown as the dotted curves in Fig.\ \ref{Fig:ucomp}).  Even though
the asymmetry in the zonal circulation is small, the Voyager results
are clearly not halfway between the current and reversed states.  They
are instead in very good agreement with the current drift rate
observations, suggesting that the asymmetry in the zonal circulation
may be a permanent one.  Although there is no known physical basis for
such a permanent asymmetry, it cannot be ruled out, especially given
the substantial and non-seasonally reversing asymmetries in the
circulations of Jupiter and Saturn.

\section{The Behavior of Long-lived Discrete Features}

\cite{Kark1998Sci} tracked seven cloud features in near-IR HST NICMOS
images captured during July and October 1997, and noted that during
the entire time span he found no evidence for appearance or
disappearance of any of features.  \cite{Hammel2005winds} and
\cite{Sro2005dyn} showed that not all Uranus cloud features share this
persistence. However, \cite{Sro2005dyn} did find seven long-lived
cloud features during the 2003-2004 time period, most with estimated
lifetimes longer than a month, one longer than a year, and one that
appeared to persist for a decade or longer. We also found long-lived
features in our 2007 data.

\subsection{Tracking of long-lived cloud features}\label{Sec:lifetime}

\begin{figure}\centering
\hspace{-0.2in}\includegraphics[width=3.55in]{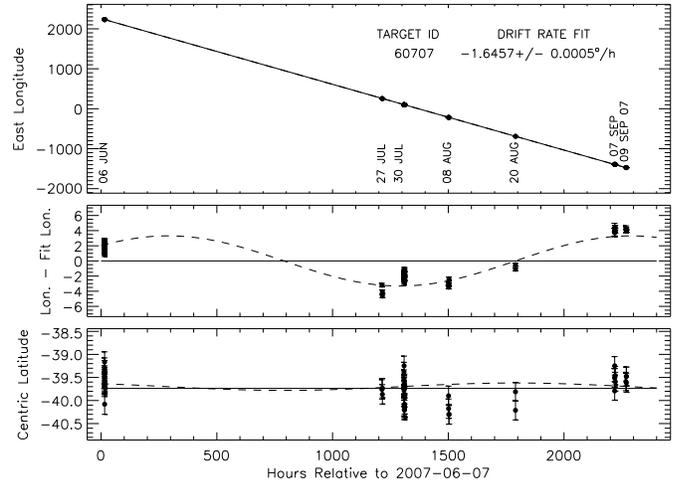}\par
\caption{Cloud target 60707 longitude vs. time (top), longitude relative to a constant
drift model (middle) and latitude vs. time (bottom).  The longitude deviations are
roughly consistent with an oscillation of longitude amplitude (0-peak) of $\sim$3\degx 
and a period of $\sim$2000 hours (83 days).}
\label{Fig:60707}
\end{figure}

Among the 28 cloud features stable enough to track over more than a
single rotation of Uranus, seven were tracked for more than 1000
hours, one of which was tracked for more than 2250 hours.  The latter
feature (labeled 60707 in Fig.\ \ref{Fig:2x3jul27}) is a
relatively subtle brightening at the edge of the southern bright band,
best seen in the high-pass filtered image for 27 July.  It can also be
seen near the central meridian in the 31 July image in Fig.\
\ref{Fig:2x3jul30}, and near the right limb in Fig.\
\ref{Fig:2x2sep7}.  A plot of longitude and latitude vs. time
(Fig. \ref{Fig:60707}) indicates both short-term and long-term
variability in the position of the feature.  Most of the short-term
variability is due to the fuzzy longitudinally extended nature of the
feature, which creates uncertainty in defining its exact location in
longitude.  The long-term variations may be due to wave
motion. Although the existence of a stable oscillation cannot be
confirmed due to sampling limitations, the deviations from a constant
longitudinal drift are consistent with a modulation amplitude (0 to
peak) of $\sim$3\deg and a period of $\sim$83 days. That period is
much shorter than the period of $\sim$1000 days found for the large
feature near 34\deg S \citep{Sro2005dyn} or the periods of 448 or 753
days suggested for the brightest ever Uranus cloud feature found at
30\deg N \citep{Sro2007bright}.  If the longitude oscillation is
attributed to a small latitudinal oscillation and an associated
variation in drift rate due to the latitudinal gradient in the zonal
atmospheric flow, then the amplitude of the latitudinal oscillation
would be only $\sim$0.08\degx , and thus not observable. (Equations
relating latitudinal and longitudinal oscillations are given in Sec.\
\ref{osc30}.)

\subsection{Reacquisition of a major northern bright spot (and dark spot?)} 

\begin{figure}\centering
\includegraphics[width=1.6in]{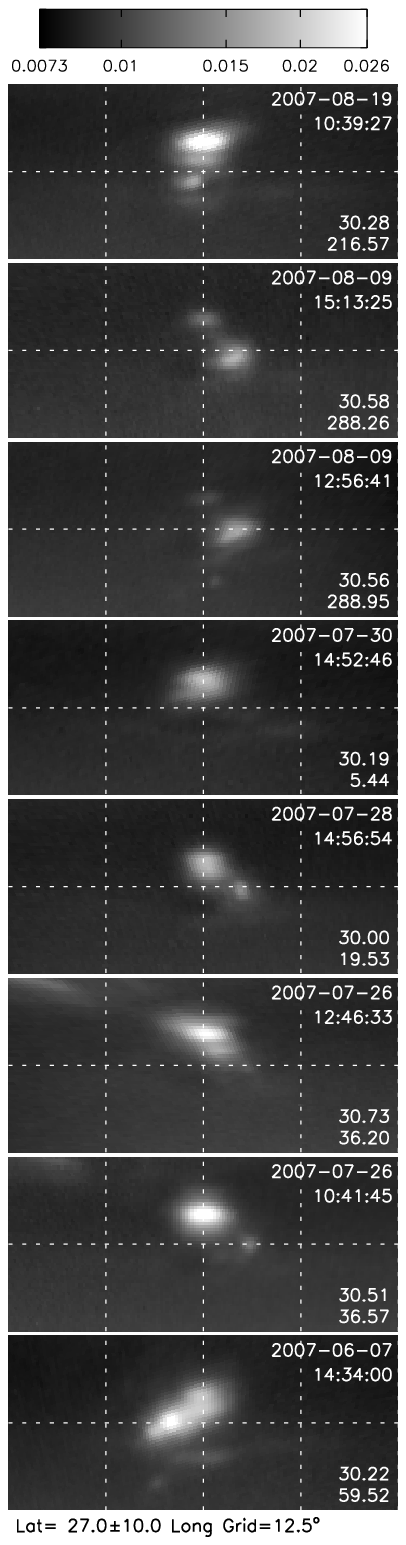}
\includegraphics[width=1.6in]{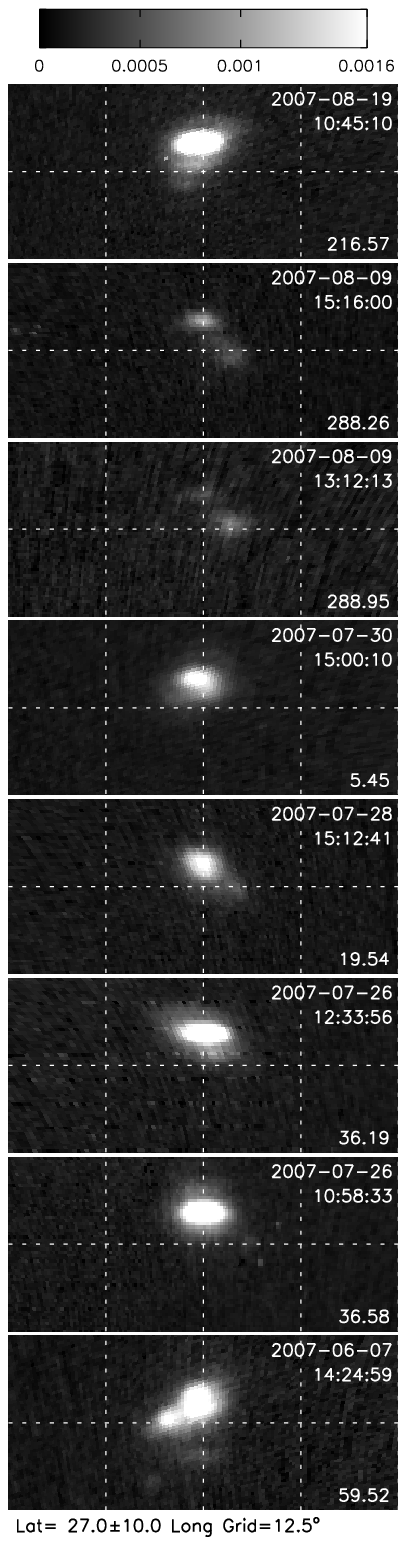}
\caption{Rectilinear projections of the Bright Complex feature using H
images (left) and K$'$ images (right).  The map display following the
first is offset by the mean drift rate of target 60720 (-0.3202\degx
E/h) times the time difference relative to the first image. All
projections were enhanced to the same I/F range to show the apparent
brightness changes, as well as changes in feature location. The H
images were logarithmically enhanced to better show dim and bright
features with the same enhancement.  The grey bars at the top indicate
the I/F range displayed in the images.
Vertical
variations of the feature components are indicated by varying ratios
of K$'$ to H. }
\label{Fig:30Nstrips}
\end{figure}

The next longest tracking interval (1751 h) was achieved for the
target labeled 60720 (visible in Figs. 2-4).  This is part of a
complex of features near 30\deg N that probably persists for much
longer periods than any individual element and is likely related to an
especially prominent bright spot near 30\deg N latitude, which in
August 2005 became the brightest cloud feature ever observed on
Uranus. That feature was also found to exhibit two superimposed
oscillations, a slow 445-day oscillation in longitude relative to its
mean drift rate and a rapid 0.7-day inertial oscillation
\citep{Sro2007bright}. A similar feature (possibly the same feature in
a different state of activity), was seen in 2007.  A sequence of
images of this feature, shown in Fig.\ \ref{Fig:30Nstrips}, suggests
that the bright feature is possibly a companion to a dark feature.
Other bright companions seem to be present as well, with varying
relative locations and brightness values. In two cases, in the bottom
row and 2nd from the top row in Fig.\ \ref{Fig:30Nstrips}, the
brightest feature in the H image is different from the brightest
feature in the K$'$ image, which is a more sensitive indicator of
cloud altitude.  Temporal variations in altitude are also present as
indicated by varying relative brightness in K$'$ images.

\begin{figure*}\centering
\includegraphics[width=1.95in]{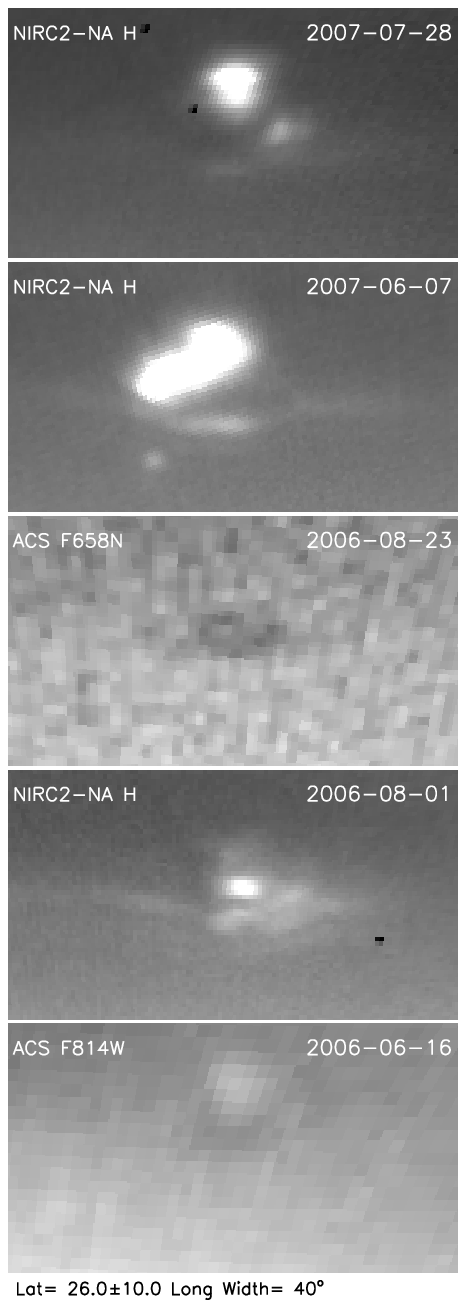}
\includegraphics[width=2.65in]{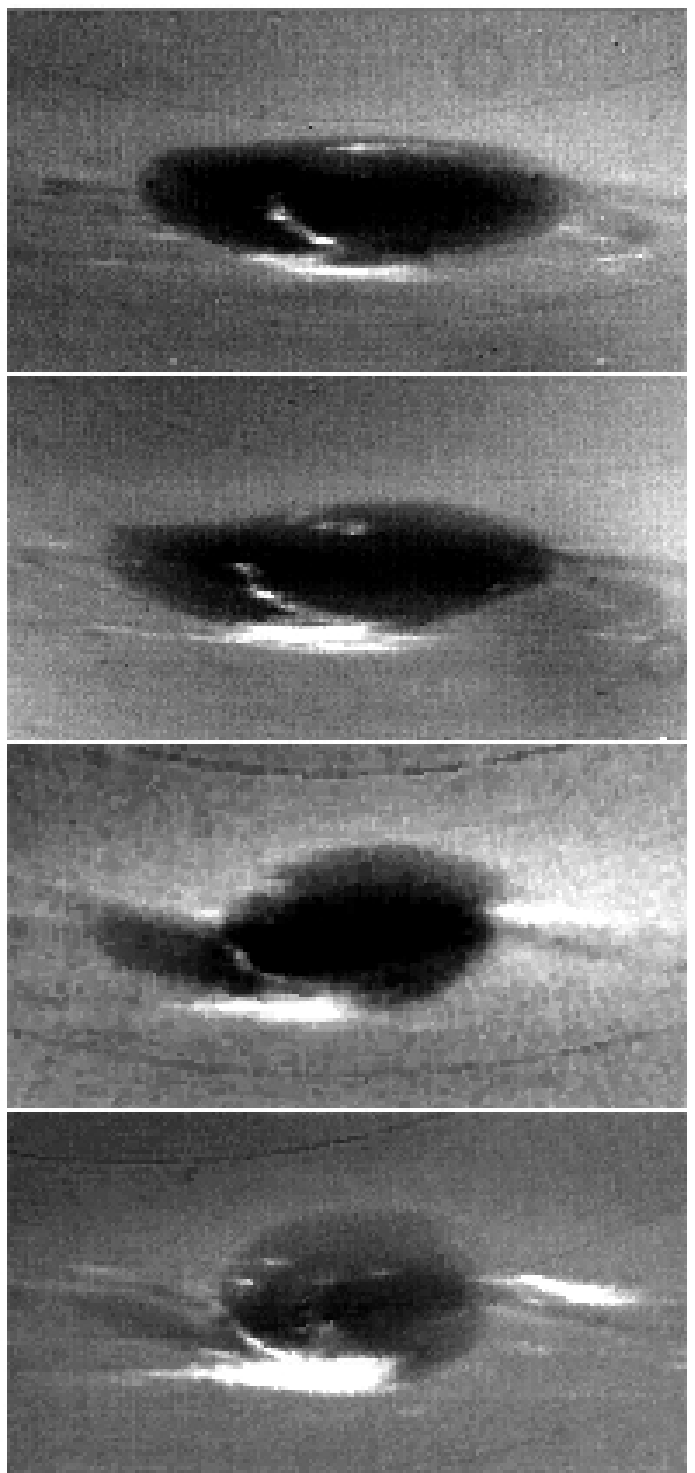}\par
\caption{ Uranus Dark Spot images (left column) in 2006 (lower three
images) and possible sightings in 2007 (upper two images). The
discovery image is the central image in the left column, taken by HST
ACS Camera with the F658N filter \citep{Hammel2008spot}.  The Keck II
NIRC2 sightings in 2006 and 2007 show bright companions similar to
those of Neptune's GDS (right column, adapted from Fig. 3 of
\cite{Sro1993Icar}). The GDS panels are centered near 15\deg N and
span about 60\deg in longitude and 35\deg in latitude.}
\label{Fig:darkspot}
\end{figure*}

This northern complex appears to be related to the first dark spot
discovered on Uranus.  While Neptune is famous for its Great Dark
Spots, until recently there had been no comparable feature observed on
Uranus.  The first such Uranus Dark Spot, discovered with HST in 2006
\citep{Hammel2008spot} and shown in Fig.\ \ref{Fig:darkspot},
exhibited a surprisingly large contrast in the red and was faintly
visible in Keck near-IR images in 2006.  A compact dark feature was
also observed on 7 June 2007 in Keck images (also in Fig.\
\ref{Fig:darkspot}), along with bright companion clouds much like
those observed for Neptune's Great Dark Spot in 1989 (right column of
Fig.\ \ref{Fig:darkspot}).  Both HST and Keck features are of similar
size, about 9-10\deg in longitude and about 3.5\deg in latitude, and
at similar latitudes of 25-26\degx N.  The Keck feature in 2007 is
just south of 60720 in Fig.\ \ref{Fig:2x3jul27} and is also visible in
Fig.\ \ref{Fig:30Nstrips}.

\subsection{Possible oscillation of the 30\deg N complex}\label{osc30} 

\begin{figure}\centering
\hspace{-0.2in}\includegraphics[width=3.55in]{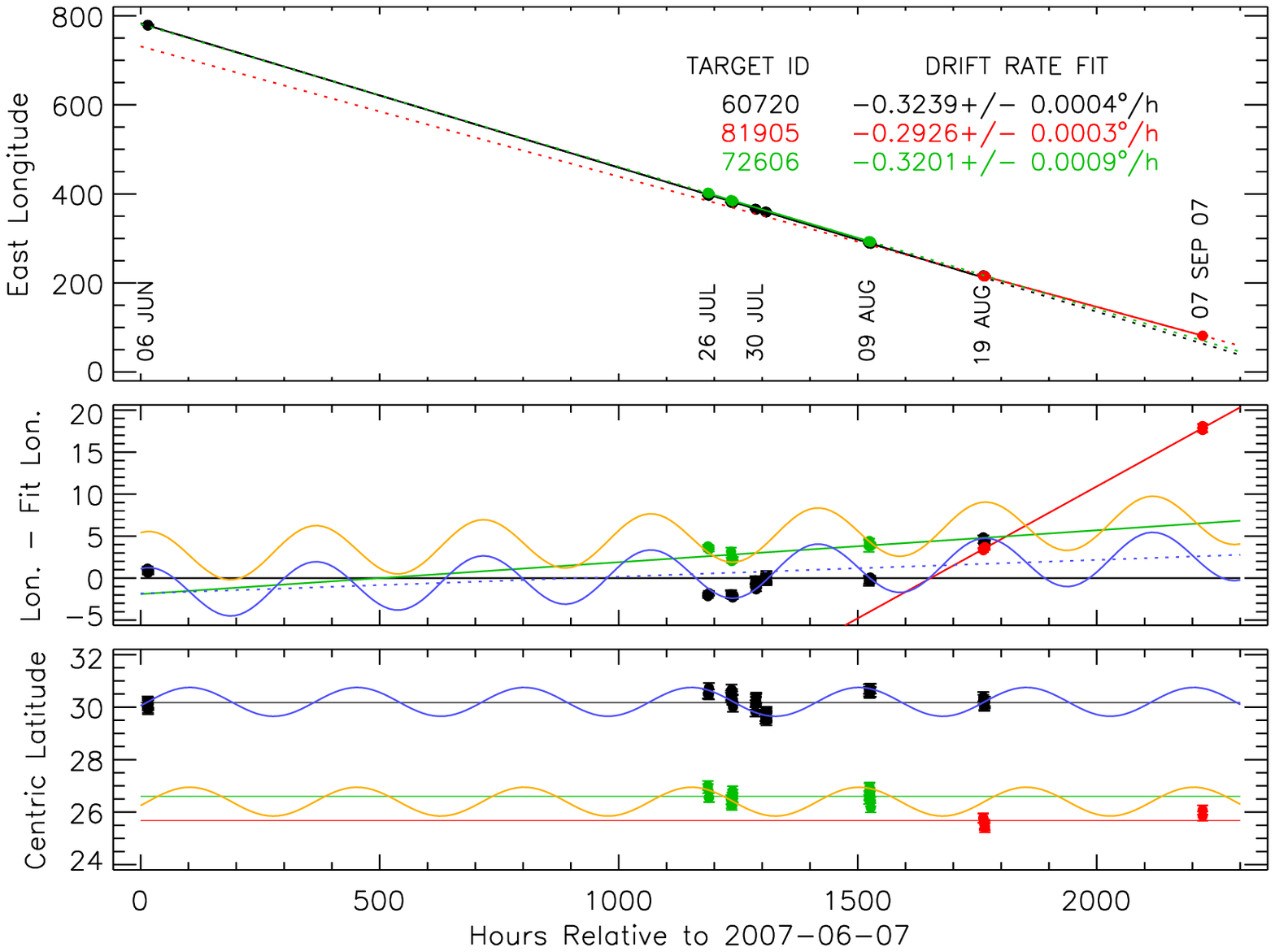}
\caption{Latitude (bottom), longitude (top), and relative longitude
(middle) for three bright cloud features associated with the BS30N
complex. The feature tracked over the longest period (60720) provides
the mean drift reference for the middle plot.  The blue sine curves
fit latitude and longitude variations for 60720 assuming drift rate
varies with latitude in the same way as the mean zonal wind. Short
term variations are consistent with a period of $\sim$350 hours and
a zero-to-peak latitude amplitude of $\sim$0.6\degx. The
orange curves are displaced copies of the blue curves, for comparison
with 72606 measurements.  The fits for 81905 (red) indicate behavioral
changes between 19 August and 7 September (see text).}
\label{Fig:BS30Nfits}
\end{figure}

The possibly oscillatory behavior of the bright/dark complex is
illustrated in Fig.\ \ref{Fig:BS30Nfits}, which displays longitude and
latitude observations for three bright cloud features associated with
the 30\deg N complex.  Tracking the motion of the group of features is
complicated by substantial changes in morphology.  The most stable of
the various elements is the feature labeled 60720. This feature
exhibits a clear variation in latitude during July 26-31, which is
accompanied by a variation in longitude relative to its mean drift
longitude.  The best fit constant drift longitude is indicated by the
solid black curve in the top panel, which becomes a horizontal line in
the relative longitude plot in the middle panel.  These two sets of
variations are consistent with a simple model in which the latitude of
the feature oscillates in time and acquires a drift rate that matches
the zonal wind profile as it moves in latitude.  For small
oscillations we can assume a linear gradient in the zonal wind
profile.  Following \cite{Sro2007bright}, except starting with the
latitude modulation, we obtain the coupled equations
\begin{eqnarray}
  \phi(t) = \phi_{\mathrm{ref}} + a \sin(\frac{2\pi}{T}(t-t_0-t_1))\label{Eq:couple1}\\
  \lambda(t) = \lambda_1 + d_1 \times (t-t_0) - a \frac{T}{2\pi} D'
  \cos(\frac{2\pi}{T}(t-t_0-t_1))\label{Eq:couple2}
\end{eqnarray}
where, $\phi$ is latitude, $a$ is the amplitude of the latitudinal
oscillation, $T$ is the period, $t_0$ is the time reference, $t_1$ is
the time delay of the oscillation, $\lambda$ denotes
longitude, $D'$ is the gradient of the zonal wind drift (longitude
change per unit time) with respect to latitude, evaluated at the
reference latitude $\phi_\mathrm{ref}$, $d_1$ is the zonal drift rate
at the midpoint of the oscillation, and $\lambda_0$ is the initial
longitude. Because the drift rate may be controlled by a circulation
feature at a slightly different latitude than the observed feature,
and because we don't really know the zonal wind itself with sufficient
accuracy, $d_1$ is determined independently rather than just
evaluating the zonal wind profile at $\phi_\mathrm{ref}$.
   
The blue oscillatory curves in Fig.\ \ref{Fig:BS30Nfits}, which fit
the observed longitude and latitude variations of cloud 60720
reasonably well, use the model defined by Eqs. \ref{Eq:couple1} and
\ref{Eq:couple2} with parameter values $\phi_\mathrm{ref}$ = 30.2\degx
N, $a=0.55$\degx, $T=350$ h, $t_0$=1415 h, $\lambda_1=$1.0\degx,
$d_1=-0.31219$\degx/h, assuming the value of D$'$ =
-2.392\degx/day/\degx Lat given by \cite{Sro2007bright}, where here
longitudes are positive eastward. The derived value of $d_1$ is just
0.002\degx/h more than the drift rate obtained for a constant drift
model.  The gold curves, which are just offset versions of the blue
curves, provide a reasonable fit to the variations of cloud target
72606, which is offset from 60720 by 3.8\deg in latitude and 4.3\deg
in longitude.  These two features appear to move at the same drift
rate, even though widely separated in latitude, suggesting that they
are both companions to a circulation feature, such as a dark vortex,
which control both their motions.  The oscillation period of 350 h is
very different from those determined for the 30\deg N feature in 2005,
which had a rapid oscillation at an inertial period of 0.68 days and a
slow oscillation at a period most likely near 450 days, but possibly
longer \citep{Sro2007bright}, although the latitude amplitude was
comparable (0.6-0.7\degx ).  It should be noted that the best evidence
for an oscillation is near 1200 hours (see Fig.\ \ref{Fig:BS30Nfits}), where the samples are
relatively closely spaced and the feature morphology is relatively
consistent, though even there the coverage is only a small fraction of
a cycle.  The large difference in the morphology of the feature
complex between 7 June 2007 and 26 July 2007 make the linkage between
early and late observations uncertain. Thus we cannot confirm the
existence of a stable oscillation, although it does provide a
plausible possible explanation for the observed deviations in
longitude and latitude.

An especially odd result is the uncertain fate of the complex between
20 August 2007 and 7-9 September 2007.  The most consistent feature
(60720) does not appear anywhere near the predicted location in the
September images.  The feature identified as 81905, which is about
4.5\deg S of 60720, provides the most likely match to the only feature
seen anywhere near the predicted location.  However, with this
assignment the inferred drift rate (see red points and linear drift
line in Fig.\ \ref{Fig:BS30Nfits}) is very different from what
characterized the BS30N complex at prior times.  Thus it may be the
case that the vortex generating the complex of bright cloud features
changed dramatically between 20 August and 7 September, perhaps
dissipating or perhaps increasing its rate of latitudinal drift.
Given the long lifetime of this feature, and the past evidence of
great variations in the brightness of its companion features, it seems
more likely that it may be just going through a change in vertical
structure that temporarily reduces its production of companion clouds,
as was inferred for the rapid (4-day) disappearance of a bright component
of a southern feature in 2004 \citep{Hammel2005newact}.

\subsection{Dynamical changes in the southern Berg feature. }
 
\begin{figure}\centering
\includegraphics[width=1.6in]{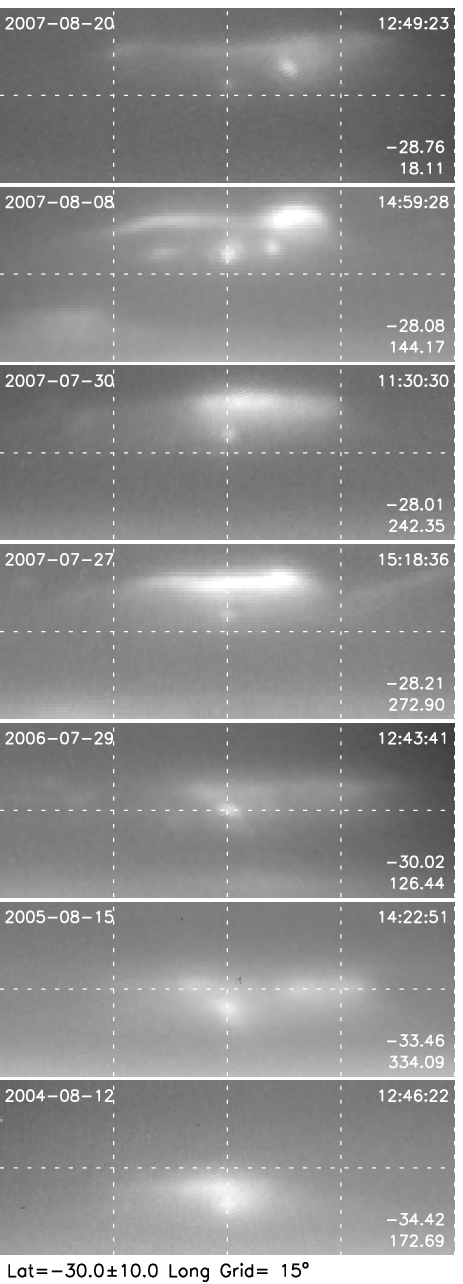}
\includegraphics[width=1.6in]{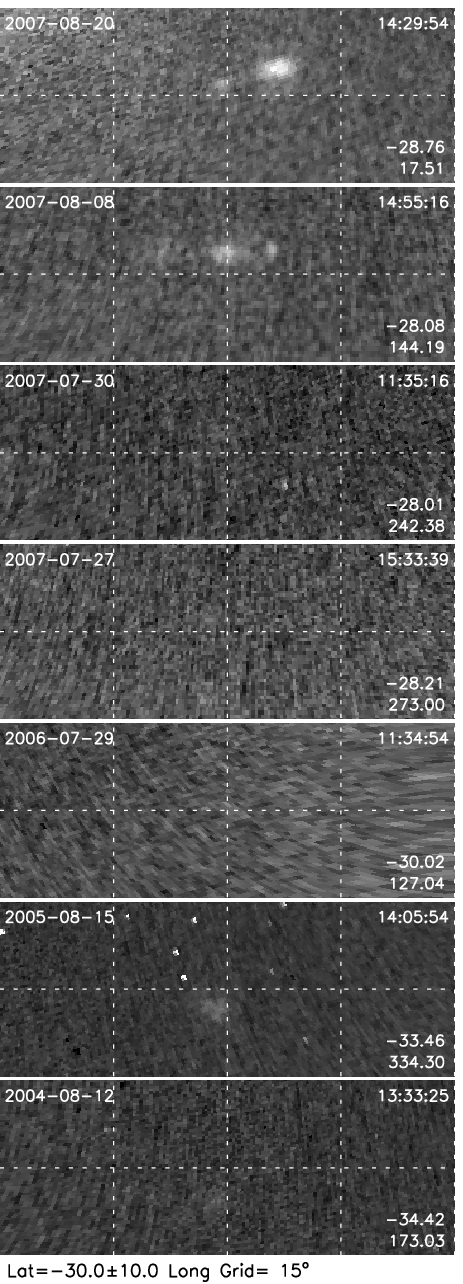}
\caption{Rectilinear projections of the Berg, previously confined
to 32\degx S - 36\degx S planetocentric, but recently moving
substantially northward and undergoing large variations in
morphology. The vertical variations in the component features are
indicated by comparison of H images (left) with the corresponding K$'$
images (right).  The central longitudes are adjusted to put the small
bright spot south of the broad streak at the center of each
projection. The actual latitudes and longitudes of the feature are
written in the lower right hand corner of image image.  To display
temporal changes in brightness, projections were similarly enhanced to
cover the same I/F range.}
\label{Fig:berg}
\end{figure}

Between 1998 and 2004 what we are now calling the Berg was the one
clearly prominent and persistent feature near 34\deg S planetocentric
latitude, which oscillated in latitude between 36\degx S and 32\degx S
with a period of $\sim$1000 days \citep{Sro2005dyn}.  This feature,
which is feature A in the results of \cite{Hammel2005winds}, has long
consisted of two components separated by about 2\deg of latitude.  The
northernmost component is generally an elongated streak, covering from
10\deg to 20\deg of longitude prior to 2005 \citep{Sro2005dyn}, but
more recently as much as 40\deg (Fig.\ \ref{Fig:berg}). A feature
morphologically similar to the 1998-2004 feature, and in the same
latitude range was found in 1994 and in 1986 Voyager observations,
suggesting a lifetime of nearly two decades.  It was thus quite
surprising to see major changes in the behavior of this feature
between 2004 and 2007, during which it began drifting northward at
about 2\deg of latitude per year (Fig.\ \ref{Fig:berglat}) and underwent large morphological
changes (Fig.\ \ref{Fig:berg}). 

The southern component of the Berg usually took the form of a single
small roughly circular bright spot (72708 in Fig.\
\ref{Fig:2x3jul27}), at least up to 30 July 2007. This is the feature
that \cite{Hammel2005newact} found to brighten significantly in 2004
K$'$ images.  However, on 8 August 2007 three small southern
components were seen (80811, 80815, and 72708, in Fig.\
\ref{Fig:2x3aug8}), and on 20 August 2007, only two were seen (80811
and 72708 in Fig.\ \ref{Fig:2x3aug19}).  Note that several small
bright spots visible in the top two H images of Fig.\ \ref{Fig:berg}
(left column) extended high enough in altitude to become visible in
K$'$ images (right column), which is very unusual for a southern cloud
feature. In fact, the sudden vertical development of the Berg in 2004
\citep{Hammel2005newact} may be dynamically related to the Berg's
subsequent northward drift. Whether the northward drift is the
beginning of an equatorial drift and subsequent dissipation, as
presumably occurred (but was unseen) for Neptune's Great Dark Spot, or
whether it will continue to exist in a new dynamical regime can only
be answered with continued observations. The Berg might also be an
analog to Neptune's second dark spot (DS2).  During the Voyager
encounter that feature was observed to oscillate $\pm$2.4\deg in
latitude and $\pm$47\deg in longitude with a 36-day period
\citep{Sro1993Icar}.  The oscillation of DS2 was never explained, nor
was its absence in any subsequent images of Neptune. Perhaps it
stopped oscillating and began an equatorial drift that ultimately
resulted in dissipation.  Thus, its fate might be better understood
from a more detailed history of the Uranus Berg.

\subsection{Motion of the Berg in 2007. }

\begin{figure}\centering
\hspace{-0.15in}\includegraphics[width=3.5in]{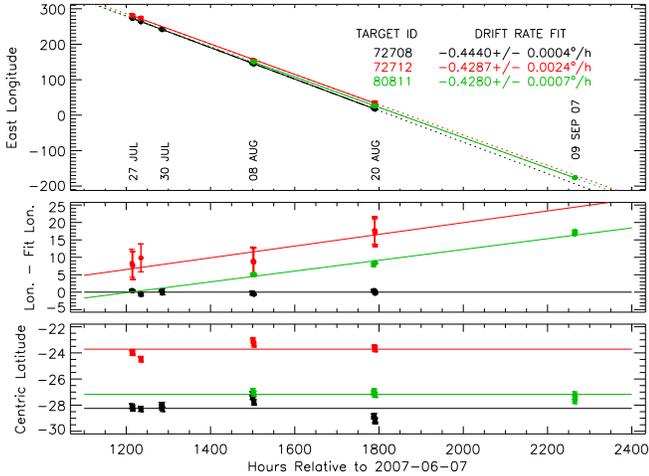}
\caption{Motion of components of the  complex southern Berg feature in 2007. 
Longitude and latitude observations for three of the longer lived
components (72708, 72712, and 80811) are shown in plots of longitude
vs. time (top), difference of the measured longitude relative to the linear
drift fit to 72708, and planetocentric latitude (bottom).}
\label{Fig:bergdrift}
\end{figure}

The motion of the Berg in 2007 is complicated to define because its
component elements undergo such dramatic morphological changes.  Each
component can only be tracked for a small fraction of the total
lifetime of the complex.  Targets that were part of the complex
included, 72708, 72712, 82007, 80811, and 80815.  The most prominent
of these was 72708, a small well-localized bright feature a few
degrees south of the main streak feature. It was tracked from 27 July
2007 to 20 August 2007, and can be seen in the left column of images
in Fig.\ \ref{Fig:2x3aug8}, including the K$'$ image, which indicates
that it reached unusually high altitudes for a southern hemisphere
cloud feature.  Two other features (80811 and 80815) also appear in
this image, and can also be seen, though with low contrast, in the
K$'$ image. Another prominent feature (72712) is a brighter and
thicker region within the elongated streak, visible in Fig.\
\ref{Fig:2x3jul27}.  This feature is not well localized in longitude
but is well localized in latitude.  Observations of 72708, 72712, and
80811 are plotted in Fig.\ \ref{Fig:bergdrift}.  These yield very
accurate drift rates, but don't all agree with each other within their
expected uncertainties. This is not too surprising, given the
possibility of companion clouds forming in the vicinity of the
generating vortex, but not always at a stable location.

Although 72708 and 72712 differ in drift rates by 0.0153\degx /h over the 300 hours
during which they were both observed, that only results in relative longitudinal
shift of 4.6\degx , which is a small fraction of the $\sim$30\deg extent of the
Berg complex. The mean latitudes of 72712 (23.72\deg S PG) and 80811 (27.17\deg S PG)
differ by 3.45\deg and would thus have a differential drift rate of 1.38\degx /h,
implying a separation by more than 400\deg of longitude over the 300 hours,
if they had moved with the mean zonal flow at their respective latitudes.  This
assumes that the zonal flow does not actually have a nearly zero latitudinal
gradient at the latitude of the feature.  However, the gradient we infer is mainly
based on features at nearby latitudes, not at exactly the same latitudes.
Unfortunately, there is
so much uncertainty in the drift rate of other features at the same latitude,
that one cannot reach a definitive conclusion regarding the true zonal
wind gradients at the Berg latitude.

There seems to be no clear latitudinal drift or oscillation during the
43 days of tracking of Berg features in 2007. Some components do move
in latitude, but the overall behavior suggests a slight maximum in mid
August of 2007.  However, we cannot conclude that the equatorial drift
is at an end because HST images obtained in September 2008 show what
appears to be the Berg at a surprisingly low latitude of $\sim$17\deg
S \citep{Sro2008DPS}, which suggests that the Berg's equatorial drift rate has
actually accelerated.
More observations in 2009 and beyond will be needed to unambiguously resolve
the latitudinal drift and ultimate fate of this interesting dynamical feature.

\section{Possible Seasonal Changes in Latitude Bands}

\subsection{Expectations}

Because of its 98\deg spin-axis inclination, Uranus has the largest
fractional seasonal variation of insolation of any
planet. Because of its long radiative time constant, its response to
this forcing should be small and phase shifted by a considerable fraction
of the 90\deg maximum. Thus, it is plausible to suggest that Uranus'
maximum north-south contrast should occur slightly before its 7
December 2007 equinox.  Leading up to equinox, the two hemispheres have
had strong differences in the albedo of deep cloud bands (Fig.\
\ref{Fig:bandrev}) and in the dynamical activity producing localized
bright and dark cloud features.  These asymmetries are likely
indicators of a seasonal response to the latitude-dependent seasonal
variations of insolation and were expected to decline before equinox.
If Uranus were only responding to the local
annual average insolation, 
we would expect its atmospheric structure to be roughly symmetric
about the equator (the expectation would be for exact symmetry except
for the variation in Uranus' orbital distance with season). Such
approximate symmetry in structure does appear in recent microwave
observations \citep{Hofstadter2004DPS}, which probe very deep levels
where the radiative time constant greatly exceeds Uranus' 84-year
orbital period. But at higher levels of the atmosphere, we have seen
hemispheric contrast changes suggesting seasonal effects. The earliest
changes were in the south polar cap, and the most recent in the bright
band structure near 45\deg S.

\subsection{The South Polar Cap}

Analysis of 1986 Voyager observations by \cite{Rages1991} inferred the
existence of a southern polar cap of relatively thick clouds. However,
the presence of single permanent south polar cap was strongly rejected
by the light curve analysis of \cite{Hammel2007var}, as well as by the
post 1994 imaging results.  The bright cap was still prominent in
1994, but declined significantly between 1994 and 2002
\citep{Rages2004Icar}, leaving primarily a narrow bright band near
45\deg S. \cite{Rages2004Icar} interpreted the changed appearance as a
decline in the optical depth of the methane cloud, placed between 1.26
and 2 bars, which resulted in better views of deeper cloud patterns at
the 4-bar level, although changes in the polar region methane mixing
ratio may have been an even more important factor \citep{Kark2008DPS}.

\subsection{The bright bands near 45\deg}   
The ``collar'' that used to mark the boundary of the south polar cap
is now visible as a bright band in images of intermediate absorption,
such as the H-band images of Figs. 2-6.  At short wavelengths, where
little absorption is present, it is overwhelmed by Rayleigh scattering,
and at strongly absorbing wavelengths it is too deep to be seen.  The
band extends from approximately 40\deg S to 48\deg S in planetocentric
coordinates (41.3\deg to 49.3\deg planetographic). 
Near-IR grism observations of Uranus in 2006 from the Keck telescope
 indicate that the bright band is associated with increased aerosol
 scattering in a cloud layer near 2 bars \citep{Sro2008grism}.  This
 cloud pressure estimate is based on methane band absorption models of
 \cite{Irwin2006ch42e} and a 1\% CH$_4$ mixing ratio.  Using a 1.6\%
 methane mixing ratio, which is closer to that estimated at the band
 latitude by \cite{Kark2009Icarus}, \cite{Irwin2007grism} also found
 increased scattering near 2 bars from an analysis of similar near-IR
 spectra of Uranus obtained at the UKIRT observatory.  On the other
 hand, \cite{Kark2009Icarus} were able to fit STIS CCD spectral images
 of the bright band using a model with locally increased aerosol
 scattering in a distributed layer between 1.2 and 2 bars. They
 argue that absorption coefficient extrapolation errors at near-IR
 wavelengths may be contributing to this pressure disagreement.

\begin{figure}\centering
\includegraphics[width=3.3in]{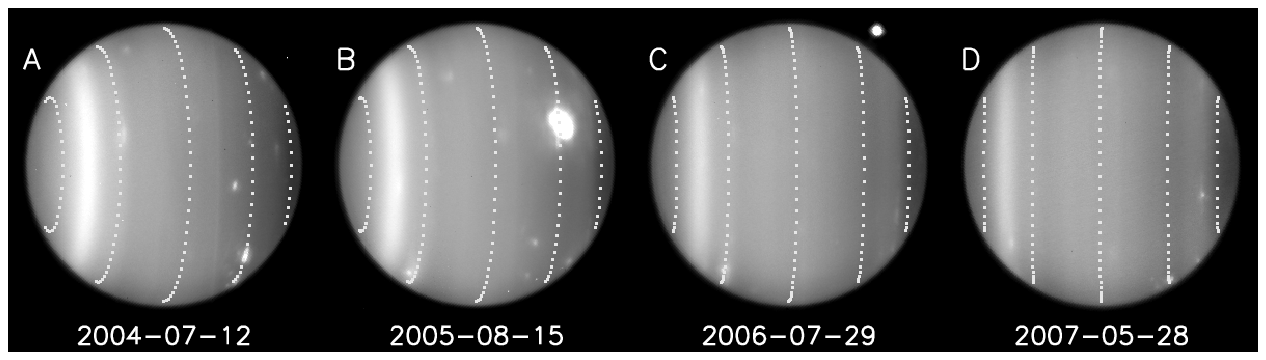}
\includegraphics[width=3.3in]{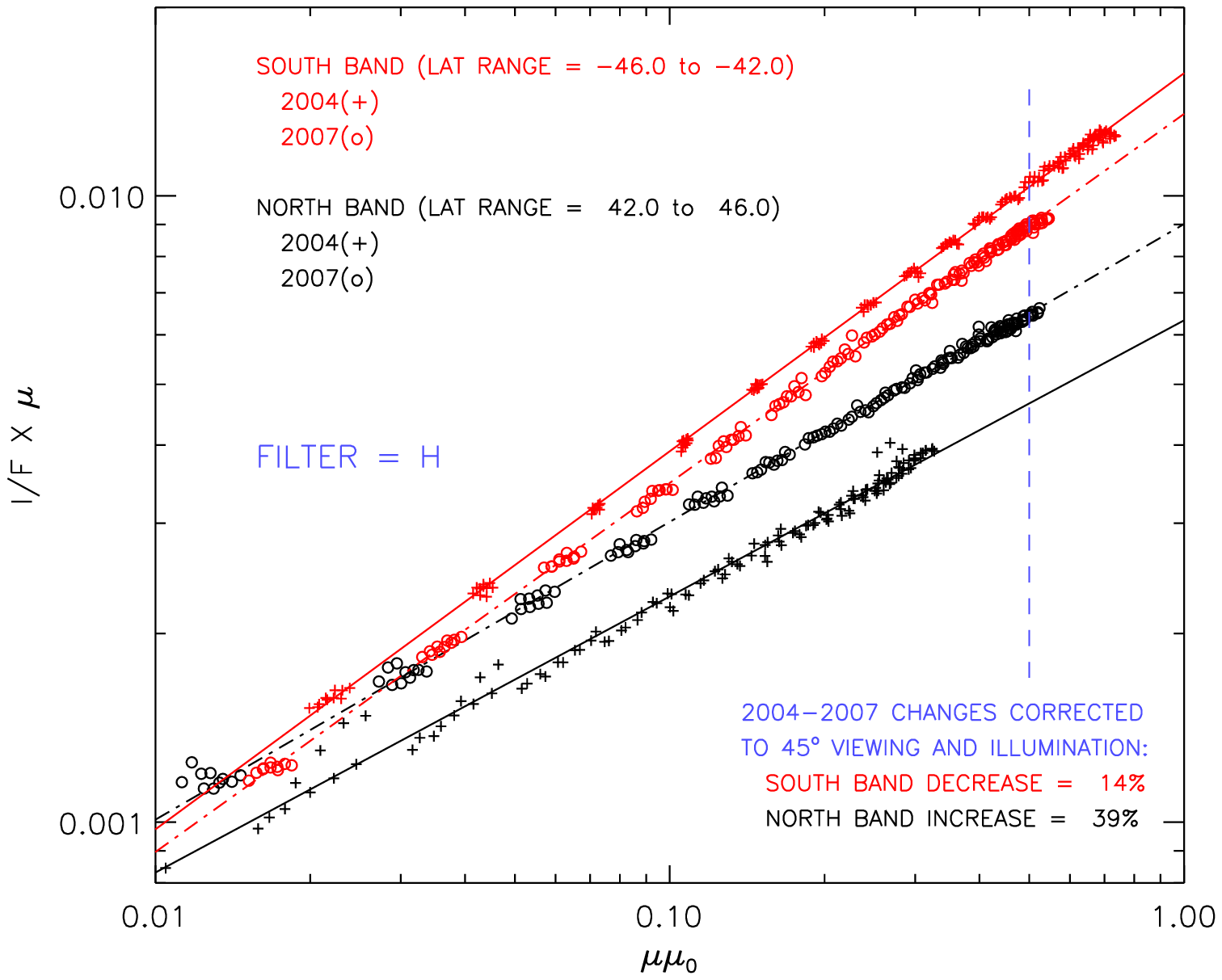}\par
\caption{Keck II NIRC2 H-band images of Uranus showing a decline of
the southern bright band and increasing brightness of a northern band,
and Minnaert plots for 2004 and 2007 showing band brightness changes
corrected for observing geometry (the brightness changes are evaluated
at $\mu=\mu_0=1/\sqrt{2}$).}
\label{Fig:bandrev}
\end{figure}

Recent HST WFPC2 and Keck observations of the bright band already
provide evidence for the expected eventual reversal in hemispheric
contrast.  This is shown in Fig.\ \ref{Fig:bandrev}, both in the
sequence of Keck H-band images and in the Minnaert plots for 2004 and
2007, which are used to correct for view angle and illumination
differences.  For each year we separately sampled reflectivity
variations within a 42\degx-46\deg S band (planetocentric coordinates)
and in what appears to be a developing northern band (using a
symmetric sampling window of 42\degx-46\deg N).  We fit these samples
to the Minnaert function $I(\mu,\mu_0)=I_0\mu_0^k\mu^{k-1}$, then
evaluated each function for the same angular conditions
($\mu=\mu_0=1/\sqrt{2}$).  Because not all of the Keck imagery has yet
been photometrically calibrated, we scaled the images to yield the
same central disk I/F value of (1.09$\pm$0.05)$\times 10^{-2}$, which is the H filter
I/F value determined by \cite{Fry2007}. The Keck H-band results reveal
a small (14\%) decline in the brightness of the south band and a
substantial (39\%) increase in brightness in the northern
band. Although absolute changes are somewhat uncertain (perhaps 5\% or
more), there is little question that the northern and southern bands
have been moving towards each other in brightness, with the northern
band clearly brightening.

\begin{figure}\centering
\includegraphics[width=3.3in]{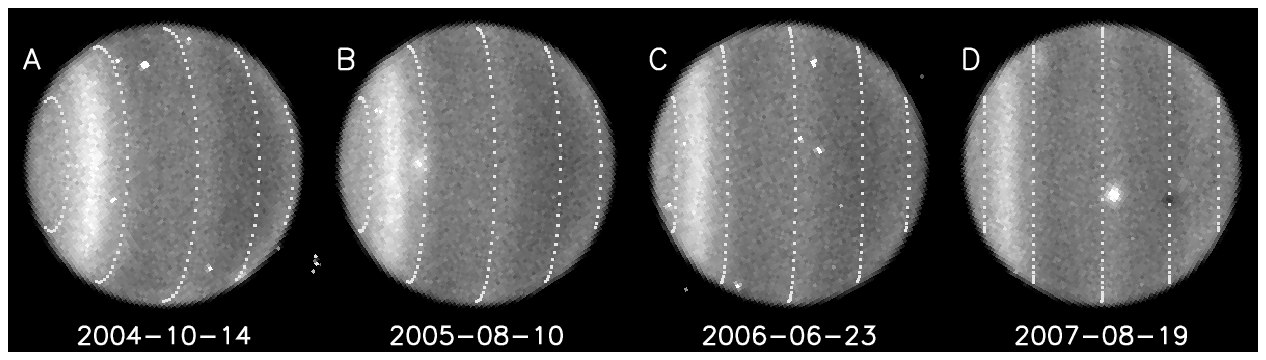}
\includegraphics[width=3.3in]{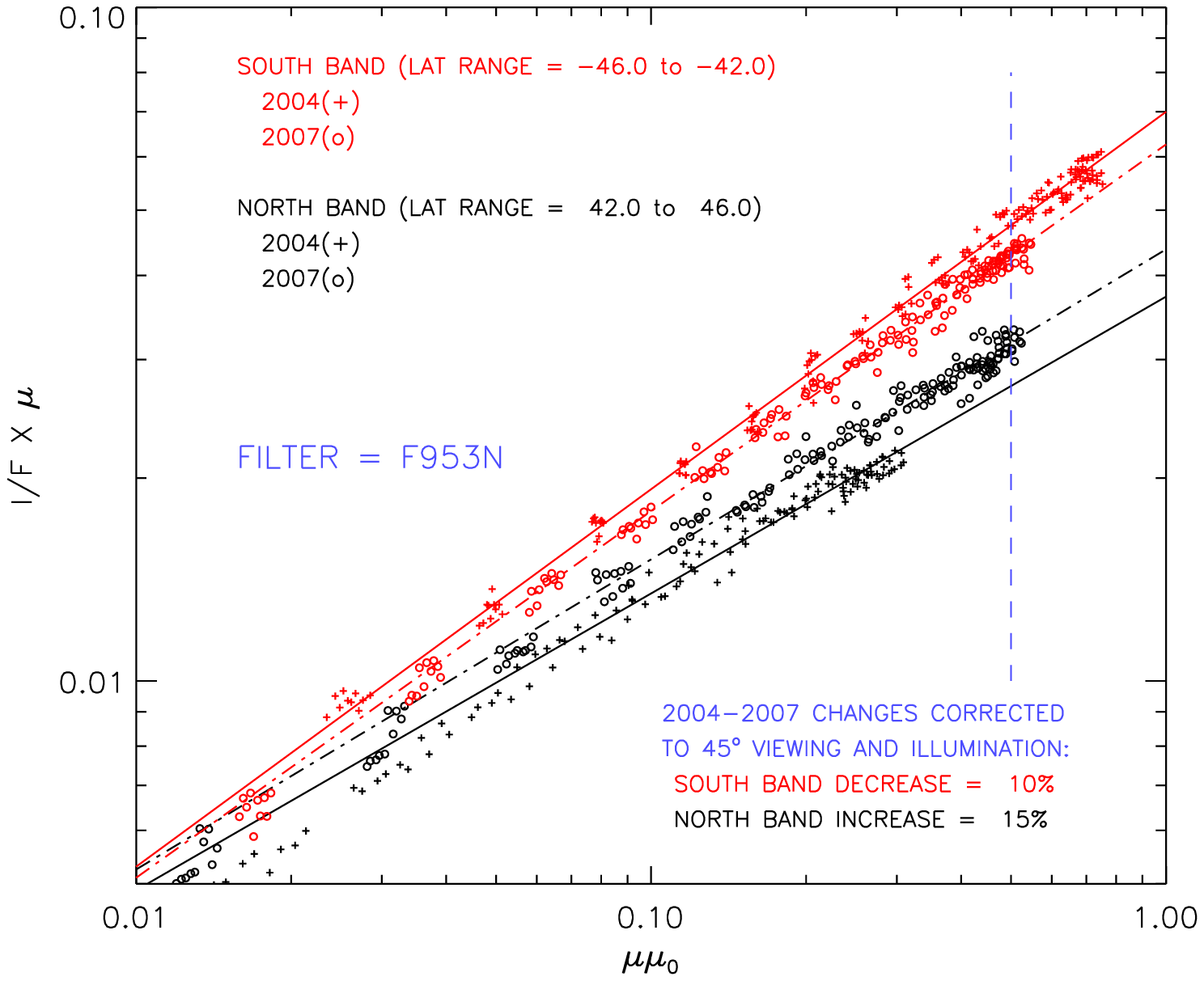}\par
\caption{HST WFPC2 images of Uranus using the F953N filter (upper)
show a decline of the southern bright band and increasing brightness
of a northern band, quantified by Minnaert plots for 2004 and 2007
showing band brightness changes corrected for observing geometry.}
\label{Fig:hstbandrev}
\end{figure}

Changes in band brightness have also been resolved by HST WFPC2 images
\citep{Rages2007DPS}, which have more accurate photometry, although
the size of the effect is much smaller at shorter wavelengths and the
effect of phase angle variations needs to be considered.
In Fig.\ \ref{Fig:hstbandrev} we display a Minnaert plot comparing HST
WFPC2 F953N images from 2004 and 2007, finding a 10\% decrease in the
southern band and a 15\% increase in the northern band; when we make
the same comparison between 2005 and 2007, which provides nearly equal
phase angles, the changes drop to 6\% and 11\% respectively, which is
approximately consistent with a linear change.  However, an analysis
of the F791W HST observations yields somewhat different results, 
indicating that imperfect
Minnaert corrections for the viewing angle differences at different
phase angles may contribute a small fraction ($\sim$1/4) of the
observed changes in HST images of Uranus between 2004 and 2007.

If we assume that the brightness changes of the two 45\deg bands
continued at the same rates observed between 2004 and 2007, they would
reach equal view-angle corrected I/F values in the F953N filter band
3.5 years after equinox (when equal solar forcing existed in both
hemispheres).  A similar calculation for the H-band suggests a phase
delay of 2.2 years.  These are surprisingly small delays compared to
the $\sim$100-year radiative time constant estimated near the 1-bar
level by \cite{Conrath1990}. Of course the temporal coverage for the
45\deg band variations is only a tiny fraction of a full cycle, and
the actual mechanism of forcing is unknown, making the meaning of the
phase delay estimates subject to considerable uncertainty.

\section{Conclusions}

HST WFPC2 imaging and near-infrared adaptive optics imaging of Uranus
by the Keck II telescope during and leading up to 2007 has revealed
new constraints on the dynamical response of the Uranus atmosphere to
seasonal forcing.  Our main conclusions are as follows:

\begin{enumerate}
\setlength{\itemsep}{-0.0in}

\item {\it Discrete cloud features.}  We found 85 features that could
be tracked well enough to yield wind errors $<$ 40 m/s, 51 of which
were tracked for intervals between 1 and 6 hours, and 28 of which
were tracked for more than a full rotation of Uranus.
Discrete clouds were found at higher northern latitudes than
previously was possible (up to 73\deg N), but otherwise the
distribution of clouds was similar to that seen in 2003-4, except for
greater numbers of clouds in the 35\degx- 45\degx N region.  No trackable
features were observed south of 42.5\deg S and few accurate winds
could be measured at low latitudes due to sparse sampling and the lack of
discrete features with sufficient contrast.

\item {\it Zonal mean circulation. }  The zonal mean circulation
measured in 2007 attained very high accuracy at 28 latitudes where
long-lived features could be tracked and medium accuracy at 10 latitudes
by binning the 1-6 hr tracking results. The extended latitude range
of our data allowed us to define for the first time a northern jet maximum near
58\deg N. We also remeasured features in 2003 images obtained by
\cite{Hammel2005winds}, correcting for a small navigation error
present in the original results and producing winds that are more
consistent with other measurements.

\item {\it Asymmetry in zonal circulation.} Uranus' zonal circulation
profile has a small asymmetry that is most apparent in the 20-40\deg
latitude range, reaching a broad peak of around 20 m/s in the
difference between north and south wind measurements at corresponding
latitudes. This can be seen in 2007 measurements, but is more firmly
established by combining measurements from 1997-2007. Symmetry
properties equatorward of 20\deg and poleward of 50\deg cannot be
assessed because of inadequate sampling.

\item {\it Temporal changes in the zonal circulation. } Uranus'
southern hemispheric circulation measured in 2007 (near equinox) does
not differ significantly from that measured in 2003-4 or from that
measured by Voyager in 1986 (near solstice).  If the observed
asymmetry near equinox were seasonally reversed, the circulation
should be midway through reversal near solstice, suggesting that
Voyager winds should be $\sim$10 m/s slower than winds near equinox.
That does not appear to be the case, suggesting that the small
asymmetry observed in the circulation profile might be a permanent
dynamical feature.  Observations over a much longer time interval will
be needed to firmly settle this issue.

\item {\it Long-lived cloud features. } Seven cloud features were
tracked for more than 1000 hours, several of which were associated
with large groups that might be considered companion clouds to vortex
circulations not directly observed.  The longest lived cloud feature
lasted for more than 2250 hours and has a non-uniform drift rate consistent
with a latitudinal oscillation with a period of 2000 hours and an amplitude
of $\sim$3\degx .

\item {\it The major northern cloud complex. }  A group of bright
features observed near 30\deg N seems to consist of companions to a
recently discovered visible Dark Spot.  The feature has non-uniform
motions consistent with coupled latitude and longitude oscillations,
with a period of 350 hours, and amplitudes of 0.55\deg in latitude and
$\sim$3\deg in longitude.  Although no individual component was
tracked throughout the observation period, the complex as a group of
variable components did exist throughout the observation period
(covering 2250 hours), and seems likely to be the same feature
associated with the 2006 Uranus Dark Spot \citep{Hammel2008spot}, and
perhaps the extremely bright cloud complex discovered first in 2005
Keck images \citep{Sro2007bright}.

\item {\it Dynamics of the Berg. } The major southern cloud feature
that oscillated for many years between planetocentric latitudes of
32\degx S and 36.5\degx S, started drifting northward in 2005,
reaching a mean latitude near 26\degx S in 2007. Remarkable
morphological changes occurred during this drift, and beginning as
early as 2004 \citep{Hammel2005newact}, some components reached
altitudes high enough to be visible in K$'$ images.  This is the only
feature observed on Uranus or Neptune that exhibited a long period of
latitudinal oscillations and subsequently transitioned to a state of
latitudinal drift.  Its subsequent fate may help us to understand the
disappearance of Neptune's second dark spot (DS2).

\item {\it Progress towards Bright Band reversal. } In the Keck II images a northern
bright band became visible in 2007, a counterpart to the bright band
extending from 40\degx-48\deg planetocentric latitude (41\degx-49\deg
planetographic) in the southern hemisphere.  Photometric analyses of
Keck and HST images show that from 2004 to 2007 the southern band has
faded by 14\% in the H band Keck II images, and by 10\% in WFPC2 953N
images, while the northern band has brightened by 39\% and 15\%
respectively.  This suggests that hemispheric symmetry at this
latitude would be achieved 2.2-3.5 years after equinox and that
bright band reflectivity has a surprisingly small
lag relative to solar forcing.

\end{enumerate}

Future observations should provide a much better definition of the
circulation at high northern latitudes and better measures of the
hemispheric contrast reversal that seems to be currently underway.  We
would expect to see a considerable brightening of the northern bright
band and the eventual formation of a northern polar cap. Hopefully,
the circulation at high northern latitudes can be well characterized
before a thickening northern polar cap obscures discrete cloud
features needed to track the winds. A breakup of the southern bright
band may permit measurement of wind speeds in this thus-far poorly
characterized region. The changes in vertical cloud structure
associated with these changes in brightness distributions on Uranus
will be investigated with the help of near-IR spectra as well as
band-pass filter imagery from both HST and Keck observatories, some of
which have already been acquired.

\section*{Acknowledgments.} \addcontentsline{toc}{section}{Acknowledgments}

We thank two anonymous reviewers for detailed reviews and constructive
suggestions, specifically for pointing out the implied short time
delay between equinox and projected bright band symmetry.  This work
is partly based on observations with the NASA/ESA Hubble Space
Telescope. Support for Program numbers 10805, 11118, 11156, 11190, and
11292 was provided by NASA through grants from the Space Telescope
Science Institute, which is operated by the Association of
Universities for Research in Astronomy, Inc., under NASA contract
No. NAS5-26555.  Support for this work was also provided by NASA
through its Planetary Atmospheres Program under grant NNG05GF00G (LAS)
and the Planetary Astronomy Program under grant NNG05G084G (LAS). HBH
acknowledges support from NASA grants NNG0GG125G, NNX06AD12G, and
NNX07A043G, as well as funding from NASA specifically for Keck
observations and reductions. IdP has been funded by NASA grant
NNX07AK70G, and by the National Science Foundation and Technology
Center for Adaptive Optics, managed by UC Santa Cruz under cooperative
agreement No. AST-9876783.  Some of the data presented herein were
obtained at the W. M. Keck Observatory, which is operated as a
scientific partnership among the California Institute of Technology,
The University of California, and NASA, and was built with financial
support of the W. M. Keck foundation.  The authors wish to recognize
and acknowledge the very significant cultural role and reverence that
the summit of Mauna Kea has always had within the indigenous Hawaiian
community.  We are most fortunate to have the opportunity to conduct
observations from this mountain.

\end{document}